Saarland University
Faculty of Natural Sciences and Technology I
Department of Computer Science

Master Thesis

# Index Search Algorithms for Databases and Modern CPUs

submitted by
Florian Gross

Supervisor
Prof. Dr. Jens Dittrich

Reviewers
Prof. Dr. Jens Dittrich
Prof. Dr. Sebastian Hack

# Statement in Lieu of an Oath
# Declaration of Consent

I hereby confirm that I have written this thesis on my own and that I have not used any other media or materials than the ones referred to in this thesis.

Ich erkläre an Eides Statt, dass ich die vorliegende Arbeit selbstständig verfasst und keine anderen als die angegebenen Quellen und Hilfsmittel verwendet habe.

I agree to make both versions of my thesis (with a passing grade) accessible to the public by having them added to the library of the Computer Science Department.

Ich bin damit einverstanden, dass meine (bestandene) Arbeit in beiden Versionen in die Bibliothek der Informatik aufgenommen und damit veröffentlicht wird.

| Saarbrücken,<br>November 15, 2010 | |
| --- | --- |
| (Datum / Date) | (Unterschrift / Signature) |

# Abstract

Over the years, many different indexing techniques and search algorithms have been proposed, including CSS-trees, CSB$^+$-trees, k-ary binary search, and fast architecture sensitive tree search. There have also been papers on how best to set the many different parameters of these index structures, such as the node size of CSB$^+$-trees.

These indices have been proposed because CPU speeds have been increasing at a dramatically higher rate than memory speeds, giving rise to the Von Neumann CPU–Memory bottleneck. To hide the long latencies caused by memory access, it has become very important to well-utilize the features of modern CPUs. In order to drive down the average number of CPU clock cycles required to execute CPU instructions, and thus increase throughput, it has become important to achieve a good utilization of CPU resources. Some of these are the data and instruction caches, and the translation lookaside buffers. But it also has become important to avoid branch misprediction penalties, and utilize vectorization provided by CPUs in the form of SIMD instructions.

While the layout of index structures has been heavily optimized for the data cache of modern CPUs, the instruction cache has been neglected so far. In this paper, we present NitroGen, a framework for utilizing code generation for speeding up index traversal in main memory database systems. By bringing together data and code, we make index structures use the dormant resource of the instruction cache. We show how to combine index compilation with previous approaches, such as binary tree search, cache-sensitive tree search, and the architecture-sensitive tree search presented by Kim et al.

# Acknowledgments

This work would not have been completed without the help and support of the many individual people who have helped me along the way. Particularly: Prof. Dr. Jens Dittrich for providing me with the opportunity of pursuing this research idea under his guidance, for advising me over the course of it, and for being the first to encourage me to get involved in research. Prof. Dr. Sebastian Hack for serving as my second reviewer, and for his valuable suggestions. To Dr. Martin Theobald for offering his advise. To Stefan Richter, and Christoph Pinkel for offering me excellent feedback on draft copies of this work. To Alekh, Jorge, Jörg, Stefan, Pascal, and Julian for helping out with bits and ends. To the many friends and my family for their constant support. Lastly, I am deeply indebted to Katrin for her understanding, patience, and encouragement throughout the long process of work that went into this thesis.

# Contents

# List of Figures

# List of Algorithms

# List of Tables

# Chapter 1

# Introduction

> "[T]here is a direct tradeoff between capacity and speed in DRAM chips, and the highest priority has been for increasing capacity. The result is that from the perspective of the processor, memory has been getting slower at a dramatic rate. This affects all computer systems, making it increasingly difficult to achieve high processor efficiencies." — Stefan Manegold et al. [MBK00]

This chapter will start out with a brief description of the problem of index search, and the different classes of algorithms involved therein.

The main focus will be on a subset of representative index search problems with some restrictions, e.g. assuming read-intensive workloads, such as the ones commonly seen in online analytical processing.

Most of the ideas presented, however, apply to index search in general and can be adapted for other types of index search problems with relative ease.

After giving a definition for index search, we will give a brief overview of modern hardware architecture, present a motivation for this work, and describe the contribution of this work.

This chapter will then discuss related work, and conclude by describing the structure of the remaining chapters.

## 1.1 On Index Search Algorithms

> "An index makes the query fast."
>
> — Markus Winand [Win10]

Just like the searching chapter of Knuth's book on the Art of Computer Programming [Knu98], most of this thesis is devoted to the study of a very fundamental search problem: how to find the data item that has been stored with a specific identification key.

This is sufficient for solving a wide assortment of search problems like finding the values of a mathematical function for some specific set of arguments, looking up the definition of a word in a dictionary, or the telephone number of a person based on that person's name in a telephone book.

In fact the only thing this view of index search requires is a set of records. Some part of each of those records should be designated as the key, used for identifying that record in the search index. It is possible to have multiple indices on the same data, keyed with different parts of the records. For example, there could be two indices for the same list of students: One of the indices allows finding students by their student ID number, and the other allows finding them based on their last name.

In this work, we will not consider finding multiple records at the same time. The algorithms given in pseudo-code will not directly support range queries. In case there is multiple matching records for the same key, the algorithms will only return one of the matching records. For most of the algorithms given here, it is however simple to extend them for range queries, or returning all of the records matching a key.

We also will not consider set queries, where all records matching a set of keys are returned from one search operation. There is some potential for follow-up work in this area — most of the index search algorithms discussed in this work are not explicitly concerned with that case. However, there is potential for performing better than just sequentially probing the index for all of the keys in the set. Current commercial database systems already know when to switch from index probing to a full table scan. There might be some potential for performing better than both of those when explicitly considering this case in index structures.

The algorithms given in pseudo-code in this work will return one specific value in case there is a match for the specified key, or the special value of *nil* in case there was no match. One can think of the value as being either some concrete part of a record (such as the definition of a word), the full record itself, or some kind of identifier that can be used for reading the full record from somewhere else. In either case, it does not change the semantics of the algorithms.

Figure 1.1 illustrates one possible classification of search algorithms:

- Linear search just linearly searches through an array of key-value data, stopping at the first match. It works on any key-value data without any kind of build-time preparation. Its run time is in the order of $O(n)$ where $n$ is the size of the key-value data.

- Hash-based search at build time uses a hashing function to distribute the $n$ tuples of the key-value data into $k$ buckets. The hashing function should yield an uniform distribution even for non-uniformly distributed key data. At search time it only needs to check the key-data in the bucket belonging to the search key. Its asymptotic run time is in the order of $O(n/k)$. For a large enough $k$ this can result in $O(1)$.

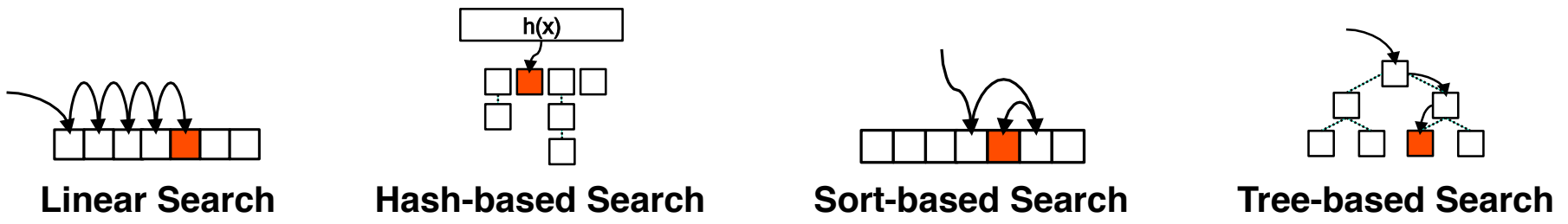

**Figure 1.1:** Classes of search algorithms [SGL09]

- Sort-based search at build time sorts the array of $n$ key-value pairs by key, saving time at search. Asymptotic run time is $O(\log n)$.
- Tree-based search creates an index tree out of the $n$ key-value pairs at build time. At run time, it traverses that tree to find the specified key. Asymptotic run time is in the order of $O(\log n)$.

All of these approaches have two operations in common: Index build-up[1], and search operations.

This work is mainly concerned with sort-based and tree-based search algorithms. There is potential for optimizing hash-based search for modern CPUs [Ros07]. We will assume that index modification will be done through a full rebuild of the index, due to focusing on read-intensive workloads, such as the ones commonly seen in online analytical processing. There is some potential for optimizing this. We will briefly talk about it later.

## 1.2 Overview of Modern Hardware Architecture

> "Reading from L1 cache is like grabbing a piece of paper from your desk (3 seconds), L2 cache is picking up a book from a nearby shelf (14 seconds), main system memory is taking a 4-minute walk down the hall to buy a Twix bar, and waiting for a hard drive seek is like leaving the building to roam the earth for one year and three months." — Gustavo Duarte [Dua08]

CPU speed has increased at a much faster pace than main memory speed. To avoid CPUs spending most of their time idly waiting for data to arrive from main memory, CPU vendors have added significant amounts of very fast cache organized into multiple layers [Int10].

[1] Which is an empty operation for linear search.

Whenever a CPU instruction needs to fetch data from main memory, it first checks the caches to see if a copy of the data has already been stored here. If that is the case, it can load the data from the cache, which is much faster than needing to go through main memory. Otherwise the data is fetched from main memory, and a copy of the data is kept in the cache, potentially evicting other data that was already in the cache.

The primary cache resources of a typical modern CPU are displayed in Figure 1.2. It can be seen that fetching data from the Level 1 and Level 2 caches carries a much lower latency than needing to read data from main memory. Level 1 Cache is split into separate data and code parts: The only way of getting things into the instruction cache is through executing code. Requests missing the Level 1 caches go to Level 2 Cache, which is unified:

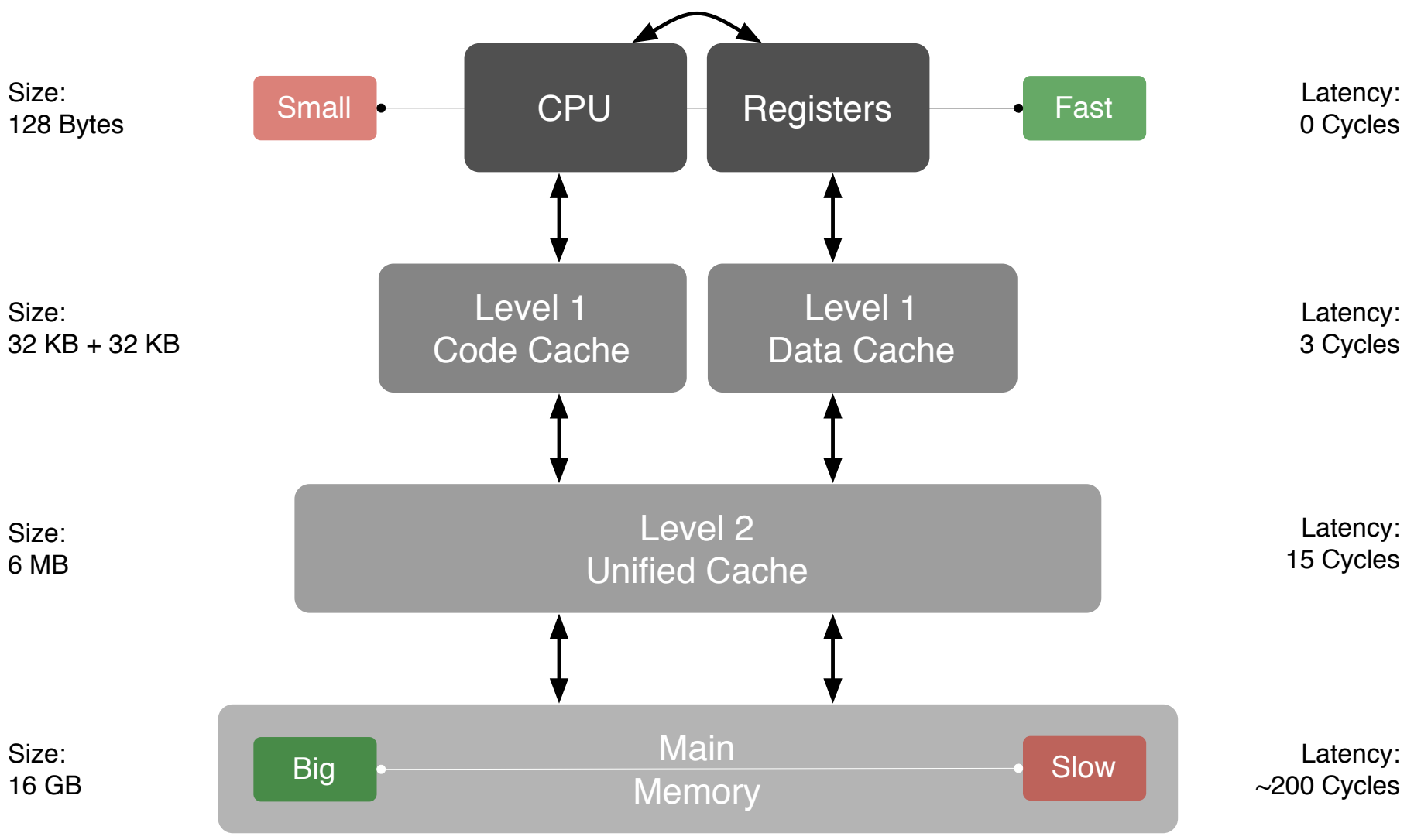

**Figure 1.2:** The cache hierarchy of modern CPUs: Storage at the top is faster, but smaller. Storage at the bottom is slower, but larger.

There is no separation of data and instructions in Level 2 Cache.

In addition to the primary cache resources that are directly used to store copies of data and code, modern CPUs also include a translation lookaside buffer: Every memory access must go through a virtual-to-physical address translation, which is in the critical path of program execution. To improve translation speed, a translation look aside buffer (TLB) is used to cache translation of most frequently accessed pages. If the translation is not found in the TLB, processor pipeline stalls until the TLB miss is served [KCS+10].

TLB misses carry the same penalty that uncached reads from main memory would carry. This means, that for a single read of data from main memory we can potentially trigger two cache misses and read requests needing to be served by main memory: One request for fetching the physical page address, and one request for fetching the actual data.

Current hardware offers support for prefetching data. The automatic prefetching done by hardware is not effective for irregular memory accesses like tree traversal, because the hardware can not predict complex access patterns. Software prefetching instructions are hard to insert for tree traversal, tree nodes far down from the current node can create a large fan out and prefetching all tree elements down wastes memory bandwidth significantly since only one of prefetches will be useful [KCS+10].

If data structures are too big to fit in caches, we should ensure that a cache line brought from the memory is fully utilized before being evicted out of caches [KCS+10].

Current hardware also offers SIMD instruction support, allowing us to use one instruction for executing the same task on multiple data elements in a single clock cycle. If it were not for SIMD, we would have to use one instruction per data element. Effectively we can save clock cycles by doing more in one cycle.

## 1.3 Motivation of this Work

Searching is the most time-consuming part of many programs, and using a good search method instead of a bad one often leads to a substantial increase in speed [Knu98].

One of the most critical database primitives is tree-structured index search, which is used for a wide range of applications where low latency and high throughput matter, such as data mining, financial analysis, scientific workloads and more [KCS+10]. Since searching is the most time-consuming part of many programs [Knu98], much time has been invested into finding good algorithms for it. The earliest search algorithm — binary search — was first mentioned by John Mauchly more than six decades ago, 25 years before the advent of relational databases [Mau46, Cod70, Knu98].

When only considering the number of comparisons needed to find a key inside an index, binary search on a sorted array is the optimal search algorithm. However, it is not true that all key comparisons come at the same cost — some comparisons are more expensive than others because of the effects of cache misses and seeks. Binary search is not optimal in reality.

Databases have traditionally wanted to avoid random access to disk at all cost due to the high seek latency of platter-based hard disk drives which vastly overshadows the computational cost of key comparison [BM72]. Index structures like $B^+$-trees reduce the number of expensive seeks required to find a key inside the index [Com79]. In recent years, as main memory continued to get much bigger and cheaper [McC09], the size of main memory available to databases has kept increasing. For most workloads, it has become possible to simply keep all data in main memory [MKB09, HAMS08, SMA+07]. Hard disk latencies have become much less important.

However, it is still not the case that all key comparisons needed for index search come at the same cost [Kam10]. If there is a cache miss for a key comparison, it can take up to $\sim 70\times$ the amount of time a key comparison without a cache miss would need.

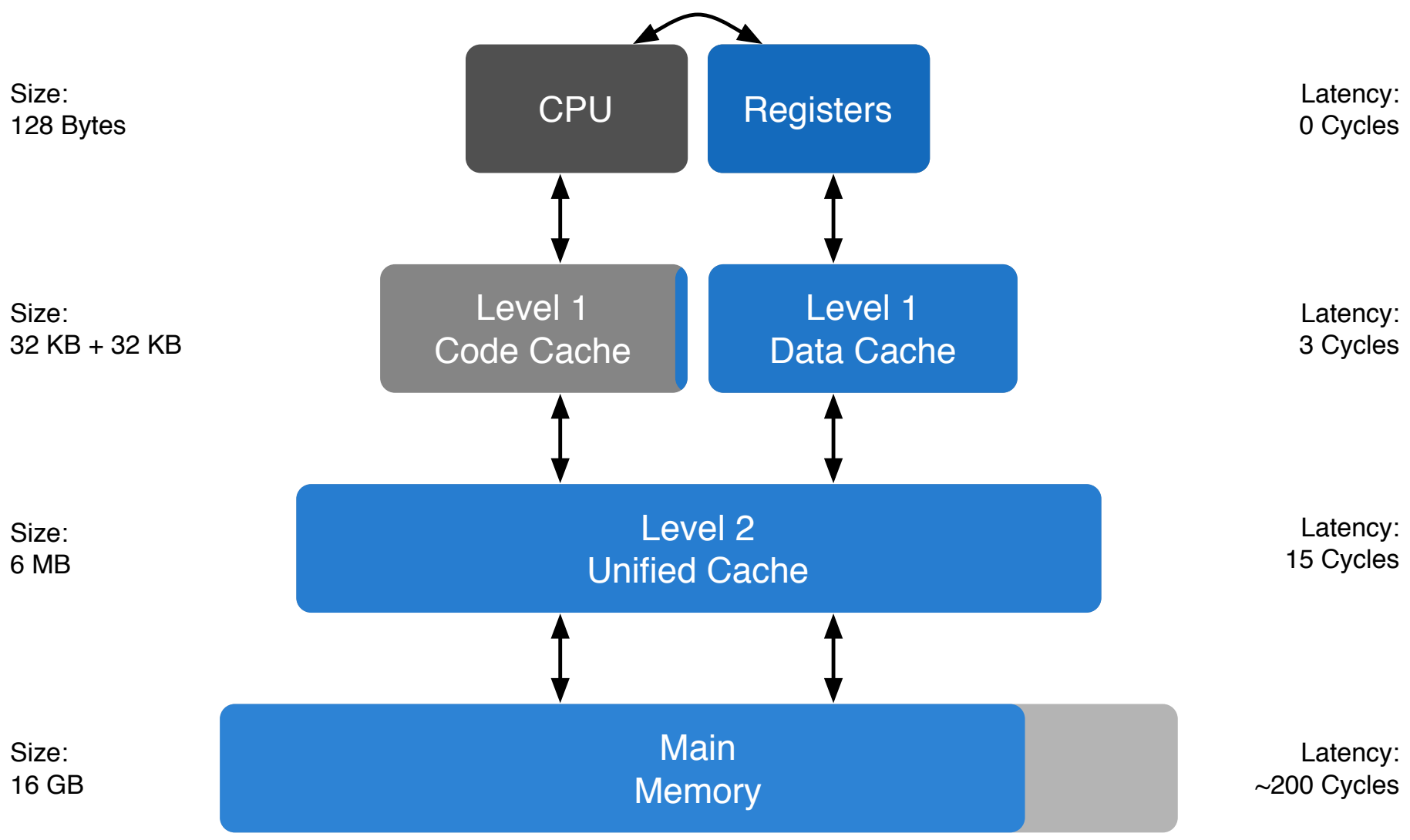

**Figure 1.3:** Utilization of CPU caches in traditional index search approaches. Blue parts are utilized by index search, gray parts are not utilized by index search. Instruction cache remains underutilized.

To avoid waiting for slow main memory, databases should well utilize these cache resource [MKB09]. Cache optimization in a main memory databases is similar to main memory optimization in disk-based ones [RR99].

Prior research has focused mainly on how best to optimize the data layout of index structures for the data cache [KCS$^+$10, RR00, RR99]. However, modern CPUs split the Level 1 cache into data cache (which is well-utilized by existing approaches) and instruction cache (which is not, see Figure 1.3).

## 1.4 Related Work

In the following we are going to briefly outline the most important pieces of work related to the ideas behind NitroGen, and index structures and algorithms.

### 1.4.1 Performance surveys

Ailamaki et al. [ADHW99] show that the optimizations found in contemporary commercial database systems did not help with taking advantage of advancements in modern CPU architecture. They report that commercial database systems spend as much as half of their execution time in stalls, not actively making use of CPU resources. Interestingly, memory stalls occur two times more frequently when using index search instead of a sequential scan in query evaluation for finding records satisfying query predicates.

Unsurprisingly, a performance evaluation done eight years later by Stonebraker et al. shows that commercial database systems still employ disk-optimized techniques for a main memory market and can be beat by up to two orders of magnitude by specialized main memory systems even in the last of their remaining application areas [SMA+07].

Manegold et al. note that a promising way of overcoming these limitations and cross the chasms between L1/L2 cache and RAM as well as utilize the processing speeds of multi-core systems is by improving the design of algorithms and data structures involved in database systems [MKB09].

### 1.4.2 Traditional index structures

The fundamental index structure used in database systems are B-trees. B-trees were originally designed as an improvement over binary tree search with the goal of avoiding slow random hard disk access as much as possible [BM72].

It achieves that goal by grouping multiple nodes of a binary search tree into bigger nodes. Whereas binary tree search executes just one comparison per tree node, B-trees can do multiple comparisons before moving onto the next node and imposing the cost of another hard disk seek. The same basic approach can be used for reducing the number of cache line misses in main memory. $B^+$-trees [Com79] increase the efficiency of B-trees by only storing key and node pointers (no value data) inside the internal nodes. This effectively lets them store more keys per node with the same node size. Leaf nodes are linked into a list to efficiently support range scans.

### 1.4.3 CPU-optimized index structures

Cache sensitive search trees [RR99] are similar to $B^+$-trees without internal node pointers. Instead, all nodes are stored in one single contiguous block of memory. In node traversal the location of the next node inside this block of memory is obtained by index arithmetic. This increases efficiency, allowing more keys to be stored per node with the same node size. Node size is tuned to match the size of a cache line, drastically reducing the number of L1/L2 cache misses. Due to all the nodes being stored in one contiguous block of memory, insertion is a costly operation. CSS-trees are best suited for read-intensive work loads such as those of decision support systems.

$CSB^+$-trees [RR00] are a compromise between the read performance of CSS-trees and the insertion performance of unmodified $B^+$-trees. By storing all children of a node of the search tree together in one contiguous block of memory, only one node pointer needs to be stored per internal node. While this does not achieve exactly the same level of efficiency as CSS-trees, $CSB^+$-trees are much easier to update than CSS-trees and still much more efficient in exploiting the cache of modern CPUs than unmodified $B^+$-trees.

Hankins and Patel [HP03] report that by tuning the node size of a $CSB^+$-

tree to be significantly bigger than cache line size they are able to reduce the number of instructions executed, TLB misses and branch mispredictions. They report that the run time effects of this overshadow the additional cache misses introduced by the bigger node size, obtaining performance of up to 1.21× the performance of CSB$^+$-trees with cache line sized nodes. We have not been able to observe a similar effect when increasing the node size of CSS-trees on present hardware architecture. This is consistent with the findings by Büttcher and Clarke [BC07].

In 1999 Intel extended the instruction set of their Pentium III processors by adding single-instruction multiple-data operations under the name of SSE. In the context of index search problems, CPUs with 128-bit registers can essentially compare four 32-bit integers in the same amount of time it would take to do a single comparison. Zhou and Ross [ZR02] propose taking advantage of this: Instead of only comparing the search key against a single separator element, they utilize SIMD to compare the search key against the separator and its neighbors. Schlegel et al. [SGL09] improve upon this: Instead of picking only one separator and splitting data into two partitions, they propose picking $k-1$ unconnected partitioning elements and splitting the elements into $k$ partitions. Due to SIMD not supporting unconnected load operations, they reorder the key array to match the order of a linearized $k$-ary search tree.

Fast architecture sensitive tree search [KCS$^+$10] combines together the ideas of CSS search trees [RR99] and $k$-ary SIMD search [SGL09]. By employing a hierarchical blocking where the data is first blocked into groups matching SIMD register width, secondly blocked into groups matching cache line size, and lastly, blocked into groups matching page size Kim et al. can utilize the speed-up of SIMD, reduce data cache misses with cache line blocking, and reduce TLB misses with page blocking which is a novelty. They also optimize their tree search algorithm for GPUs and employ compression and query pipelining to further increase throughput.

### 1.4.4 Compilation in Database Systems

Previous research by Rao et al. has used run-time compilation in database systems to reduce the overhead of passing data between the operators of a query plan [RPML06] which is similar to context threading for efficiently interpreting dynamic programming languages [BVZB05]. Earlier research proposed the use of buffer operators for reducing branch misses and thus avoid some of the overhead of passing data between operators [ZR04].

## 1.5 Description of Remaining Chapters

Chapter 2 is concerned with the most basic search techniques used in database systems. They were developed at a time when all main memory available to a database system had the size of cache resources available in today's CPUs. Consequently, those techniques are not optimized for modern hardware.

Chapter 3 then deals with search techniques which were specifically optimized for modern CPU architecture, such as CSS-Trees, CSB$^+$-Trees, k-ary Search, and FAST-Tree search.

Our contribution, NitroGen, will be covered in Chapter 4. It encompasses a detailed description of the idea of index compilation, the current status of its implementation, and future work.

Chapter 5 covers the experimental evaluation of existing indexing approaches, and our new technique of using code generation to speed up index search.

Finally, Chapter 6 concludes this work by summarizing its most important results, as well as briefly describing problems left unresolved.

# Chapter 2

# Traditional Index Structures and Search Algorithms

> "We are concerned with the process of collecting information in a computer's memory, in such a way that the information can subsequently be recovered as quickly as possible."
>
> — Donald Knuth [Knu98]

All index structures are founded on one very simple idea: When a lot of time is spent searching inside the same large set of information, then it makes a lot of sense to invest a little time upfront, and organize the information in a suitable way, such that less time needs to be spent searching later on. The goal is to organize data for fast retrieval of individual pieces of information. Less time spent searching allows us to have faster database systems.

This chapter is concerned with the most basic search techniques used in database systems. They were developed at a time when all main memory available to a database system had the size of cache resources available in today's CPUs. Consequently, these search techniques are not optimized for modern hardware. Yet, one of them is used as a part of all the more modern techniques we will focus on in Chapter 3, and the other one already uses principles very similar to those of the more modern techniques[1].

In the following, I will first briefly outline why binary search on a sorted array performs sub-optimally for searching large main memory databases on modern hardware by not optimally utilizing modern hardware resources.

I will then give a quick overview of B-trees and $B^+$-trees, which were designed to work well with large disk-based databases. I will show that they perform better than binary search on modern hardware, but still fail to optimally utilize the resources of modern hardware.

## 2.1 What's wrong with Binary Search?

Binary search operates on a linear array of sorted data, shrinking the search area by around half of the remaining keys on each search step [Mau46]. Figure 2.1 shows a sample run of binary search.

When only considering the number of comparisons needed to find a key inside an index, binary search on a sorted array is the optimal search algorithm, provably requiring the minimal possible number of comparisons. Algorithm 2.1 shows one possible implementation of binary search.

However, in order to compare the search key against an index key, we first need to fetch that index key into a CPU register. Traditionally, this fetching meant waiting for a very slow hard disk. Recently, it has become possible to keep an increasingly larger amount of data in main memory.

While it is not as slow as waiting for disks, a read from main memory is still much slower than accessing data already present in a CPU register or already available in CPU cache due to previous memory reads.

[1] The first one is binary search and the other one is $B^+$-tree search.

Modern CPUs fetch data from main memory in units of cache lines [Int10]. Reading multiple elements of data from the same cache line is much faster than reading the equivalent amount of data spread over many cache lines: The first read of data from a cache line will fetch that data to the cache if it is not already there. Subsequent reads will be able to directly read the data from cache instead of needing to go to main memory, when the temporal distance of all the reads is sufficiently low.

Because of the large distance between search keys compared in step $n$ of binary search and keys compared in step $n+1$ of binary search, one full cache line is read in step $n$, but only one of its keys is accessed. In step $n+1$ another full cache line is read, but again only one of its keys is accessed.

Figure 2.2 shows that even for the simple sample run from Figure 2.1 where only four key comparisons are made in total, there are three cache misses for a cache size of four key-value pairs. Cache misses mean that we have to read data from main memory. Fetching data from main memory into registers takes much more time than the comparison of two integer numbers. The memory loads dominate actual comparison costs. While the number of comparisons performed by binary search is optimal, the total computational time needed is clearly suboptimal when considering the effects of cache behavior in modern CPUs.

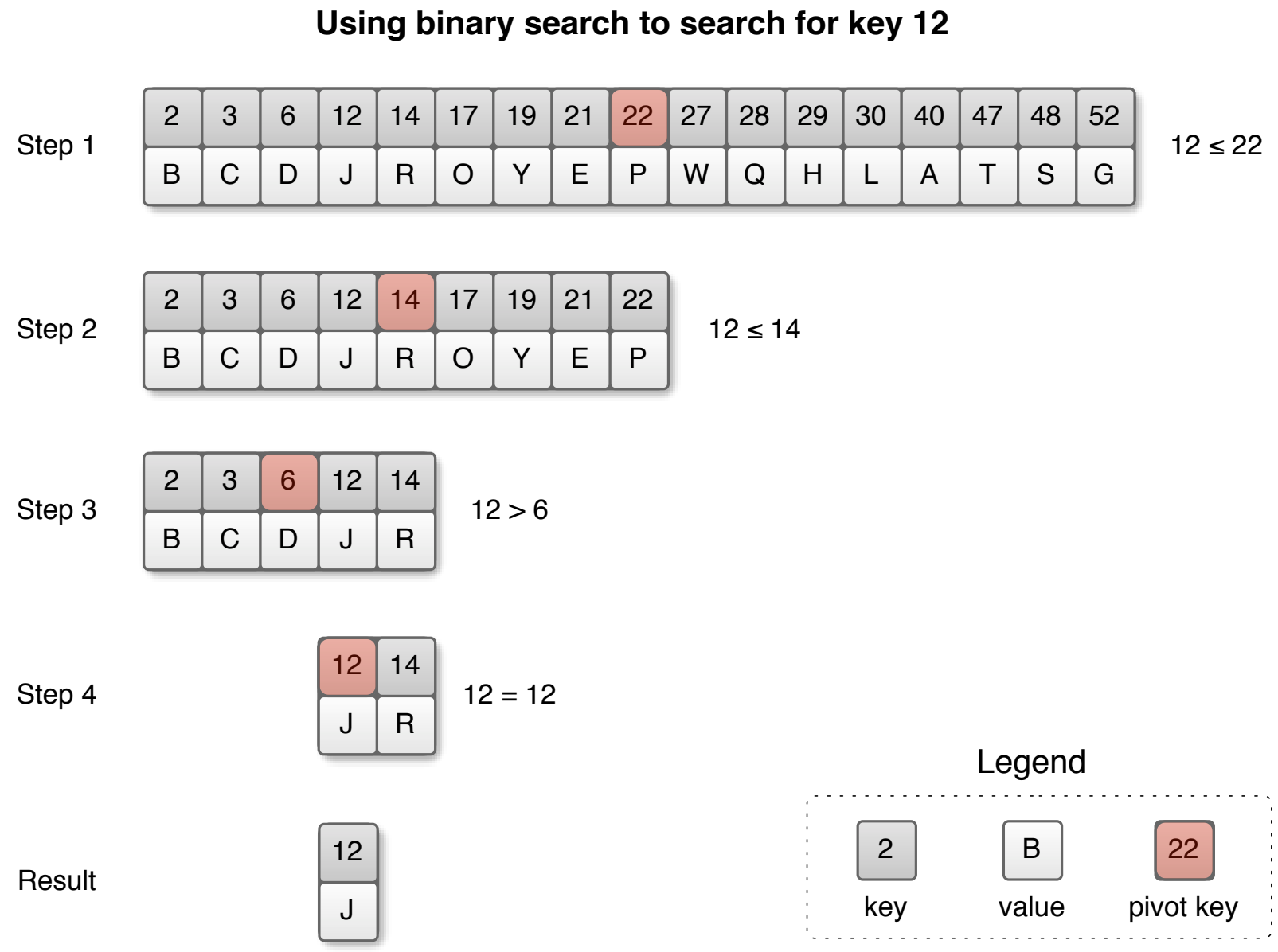

**Figure 2.1:** Binary search decreases search area by 50% after each comparison until no keys remain to be compared or a matching value is found.

Binary search does not map well to cache architecture of modern CPUs. Throughput could be much higher if subsequent reads of keys had some cache locality. This is very similar to why pure binary search is not traditionally used when reading data from disk. The same ideas used to improve disk locality can be used to improve cache locality in main memory databases.

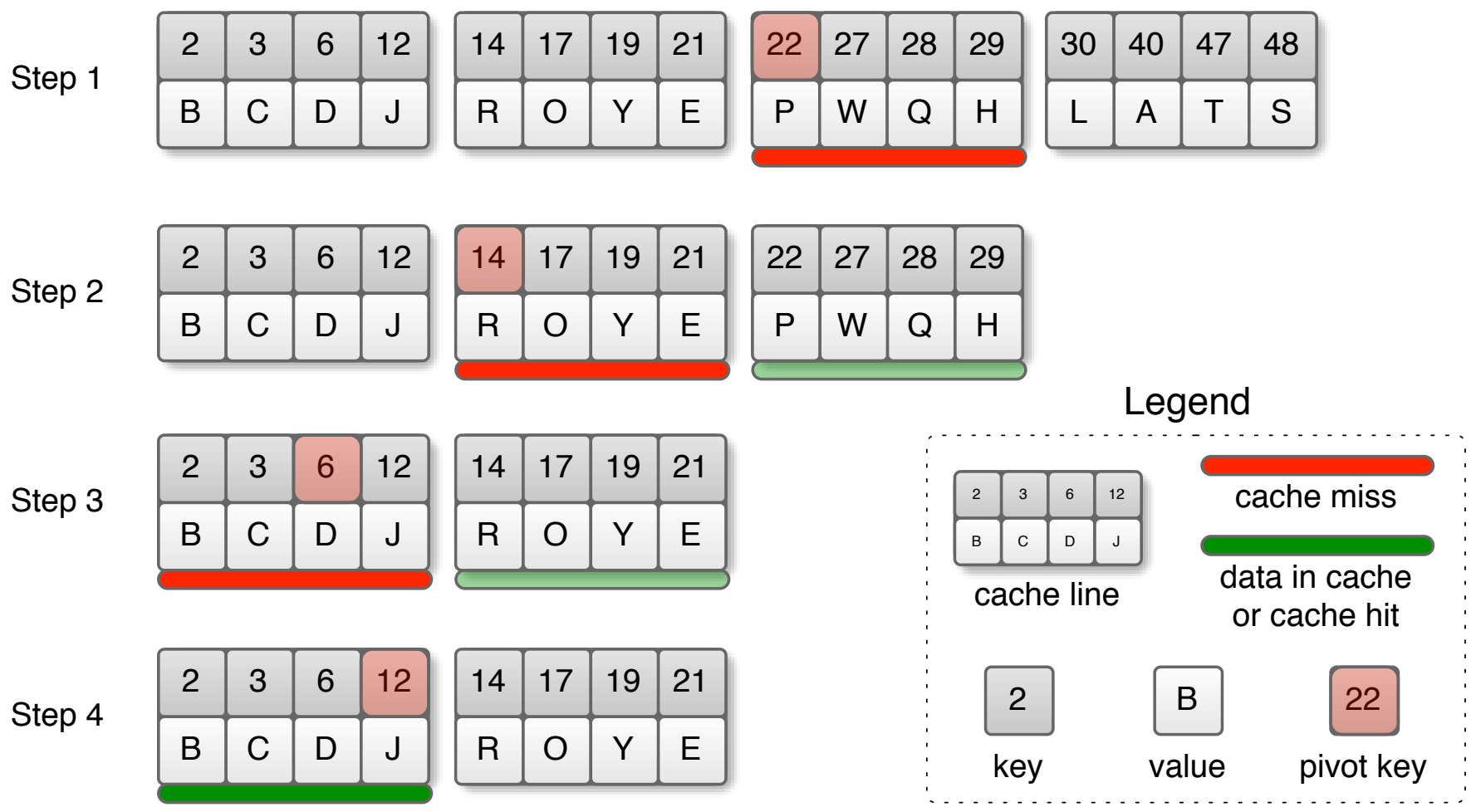

**Figure 2.2:** Cache behavior for a sample run of binary search: Three out of four key references cause slow cache misses.

```
input: Key to search for key
input: Number of data items n
input: Array of keys keys, array of values vals, both of size n
low ← 0; high ← n − 1; mid ← midKey ← nil;
while low < high do
    mid ← (low + high)/2;
    midKey ← keys[mid];
    if key > midKey then
        low ← mid + 1;
    else
        high ← mid;
    end
end
if midKey = key then
    return vals[mid];
else
    return nil;
end
```

**Algorithm 2.1**: An iterative implementation of binary search.

## 2.2 B-trees and $B^+$-trees

Instead of only fetching and comparing a single index key per search step like in binary search, B-tree[2] search compares and fetches blocks of multiple index keys[3] per search step with the motivation of reducing hard disk seeks.

The idea of grouping multiple items into nodes was originally used to reduce the number of hard disk seeks per index lookup in traditional database systems [BM72]. It can, however, be used in main memory databases to reduce the number of page misses for an index lookup as well.

$B^+$-trees [Com79] are an improvement over B-trees in that they do not

[2]The etymology of the name seems to be unknown: B-trees were developed by Bayer et al. at Boeing. They are balanced, broader than binary trees, and bushy.

[3]Bayer et al. call these blocks "pages" in their paper on B-trees [BM72].

store value data in leaf nodes, allowing better storage utilization and higher node fanout. Additionally, $B^+$-trees arrange leaf nodes as a linked list in order to allow fast handling of range queries.

Algorithm 2.2 shows a sample implementation of the $B^+$-tree search algorithm. That algorithm refers to a *binary search on key ranges*: The idea behind this is to perform a binary search on the keys of the node using the $\leq$ operator to compare keys, instead of using the $=$ operator: If $key \leq node.keys[0]$, we branch to the first child node. If $node.keys[0] \leq key \leq node.keys[1]$, we branch to the second child node etc. If the key is larger than the last of the node keys, we branch to the last child node[4].

In addition to storing keys in internal nodes, $B^+$-trees also need to store one pointer per child node. This was not necessary in binary search due to using one contiguous array of main memory: The index of the next key

[4]This is a common operation in search trees, but it appears that it does not have a proper name so far.

**input**: Key to search for *key*
**input**: Root node of tree *root*

$node \leftarrow root$;

**while** *node* is not a leaf node **do**
    $childIdx \leftarrow$ perform *binary search on key ranges* for *key* in keys of *node*;
    $node \leftarrow node.children[childIdx]$;
**end**

$keyIdx \leftarrow$ perform binary search for *key* in keys of *node*;

**if** $keyIdx \neq nil$ **then**
    **return** $node.values[keyIdx]$;
**else**
    **return** *nil*;
**end**

**Algorithm 2.2**: An implementation of the $B^+$-tree search algorithm.

to compare could simply be computed from the index of the previously compared key. Figure 2.3 shows the logical and physical layouts of B$^+$-trees and B-trees. Here each page node has three levels of binary nodes, resulting in seven keys and eight child page node pointers per page node.

Storing one pointer per child node imposes significant space overhead in B$^+$-trees: Assuming 32-bit-width integer keys and 32-bit-width pointers, more than 50% of storage space is wasted for storing internal pointer data. As a consequence, more than 50% of cache space is not used for caching actual data.

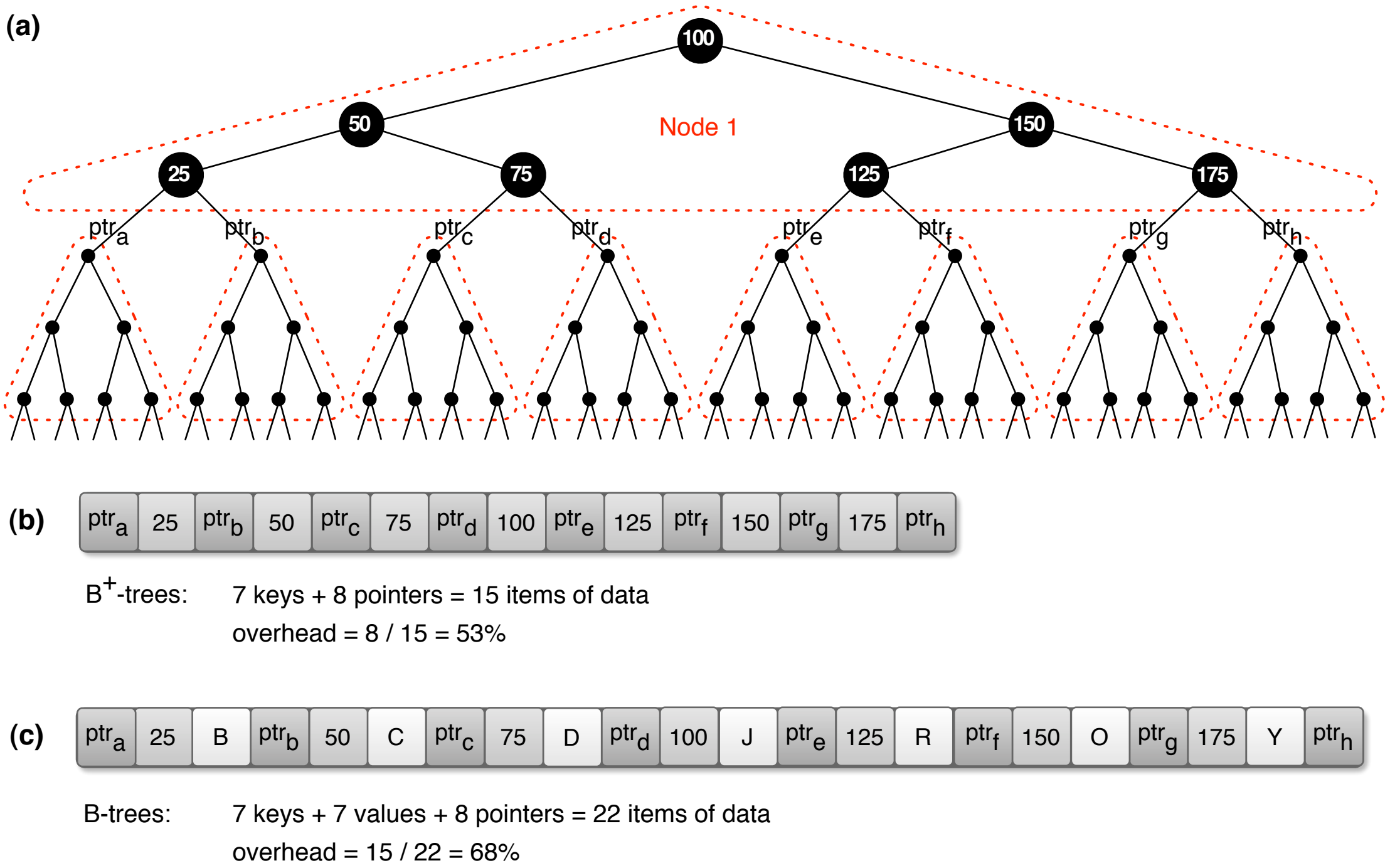

**Figure 2.3:** Logical and physical layout of B- and B$^+$-trees.
(a) Logical layout: Binary nodes are grouped into pages (red triangles).
(b) Physical layout of B$^+$-tree Node 1: More than 50% of space is wasted.
(c) Physical layout of B-tree Node 1: More than 66% of space is wasted.

# Chapter 3

# CPU-optimized Index Structures and Search Algorithms

> "Would you believe me if I claimed that an algorithm that has been on the books as "optimal" for 46 years, which has been analyzed in excruciating detail by geniuses like Knuth and taught in all computer science courses in the world, can be optimized to run 10 times faster?" — Poul-Henning Kamp [Kam10]

Taking into account the concrete properties and functionality of hardware in algorithm design leads to better performance of algorithms. Traditional search algorithms have not taken into account things such as the vast difference in speed between CPUs and main memory, which can be narrowed by good use of CPU cache resources, or the advent of SIMD technology, which allows us to perform an operation on multiple data items in the same time it would take to complete the operation for one single data item.

This chapter is concerned with search techniques that take into account the properties and functionality of modern hardware. In the following, we briefly outline how CSS-tree search [RR99] is similar to $B^+$-tree search [Com79], but with the benefit of making more efficient use of CPU caches. This comes at the cost of less flexibility in the structure of the trees, making incremental updates impossible, and requiring a full rebuild of the index structure on most changes. They are, however, still a good match for OLAP workloads.

Subsequently we will show how $CSB^+$-trees [RR00] overcome that limitation while still making much better use of CPU caches than $B^+$-trees could. They are more suitable than CSS-trees [RR99] for update-heavy workloads, such as the ones that seen in OLTP environments.

Next we will show how k-ary search [SGL09] elegantly makes use of SIMD instructions to speed up search in a sorted array, cutting in half the number of steps required for finding a key when compared to binary search when specialized for modern CPUs. However, as non-consecutive fetching of data from memory to SIMD registers is fairly expensive[1], we will see that it makes more sense to reorganize the underlying key array into a linearized tree that closely resembles the consecutive node storage already seen in CSS-tree search [RR99]. We will then show that it is desirable to have the benefits of CSS-tree search and k-ary search both at the same time to obtain optimal utilization of cache- and SIMD-resources.

We will then see how fast architecture sensitive tree search [$KCS^+10$] achieves both goals at the same time, while also adding optimal utilization of translation-lookaside buffers to further reduce the cost of memory reads in the average case.

[1] Not possible in current CPU generations without doing one separate read per data item plus costs for recombination of the data items.

## 3.1 CSS-Trees

Cache sensitive search trees (CSS-trees, [RR99]) are an implicit data structure form of $B^+$-trees [Com79]: The parent-child hierarchy of the tree structure is mapped to a linearized array.

Figure 3.1 shows the structure of an example CSS-tree. Dashed lines represent references to children that need not be physically stored as pointers. The index of the next child node to process can be calculated from the index of the current node. By encoding all the tree structure in one contiguous array, pointers to children no longer need to be stored.

We can still think of the tree as being composed of nodes, they have just been linearized into a sequential array of nodes. The size of the nodes should be optimized for two factors: *a*) matching the size of cache line as closely as possible to drive down cache accessing costs, and *b*) decreasing computational overhead costs in node processing.

By not having to store any child pointers, the amount of necessary storage shrinks to 50% of the storage needed by $B^+$-trees for storing the same amount of data[2]. The benefit of this is a better utilization of storage space, and much more importantly, better utilization of cache space. This leads to fewer cache misses and faster index lookups.

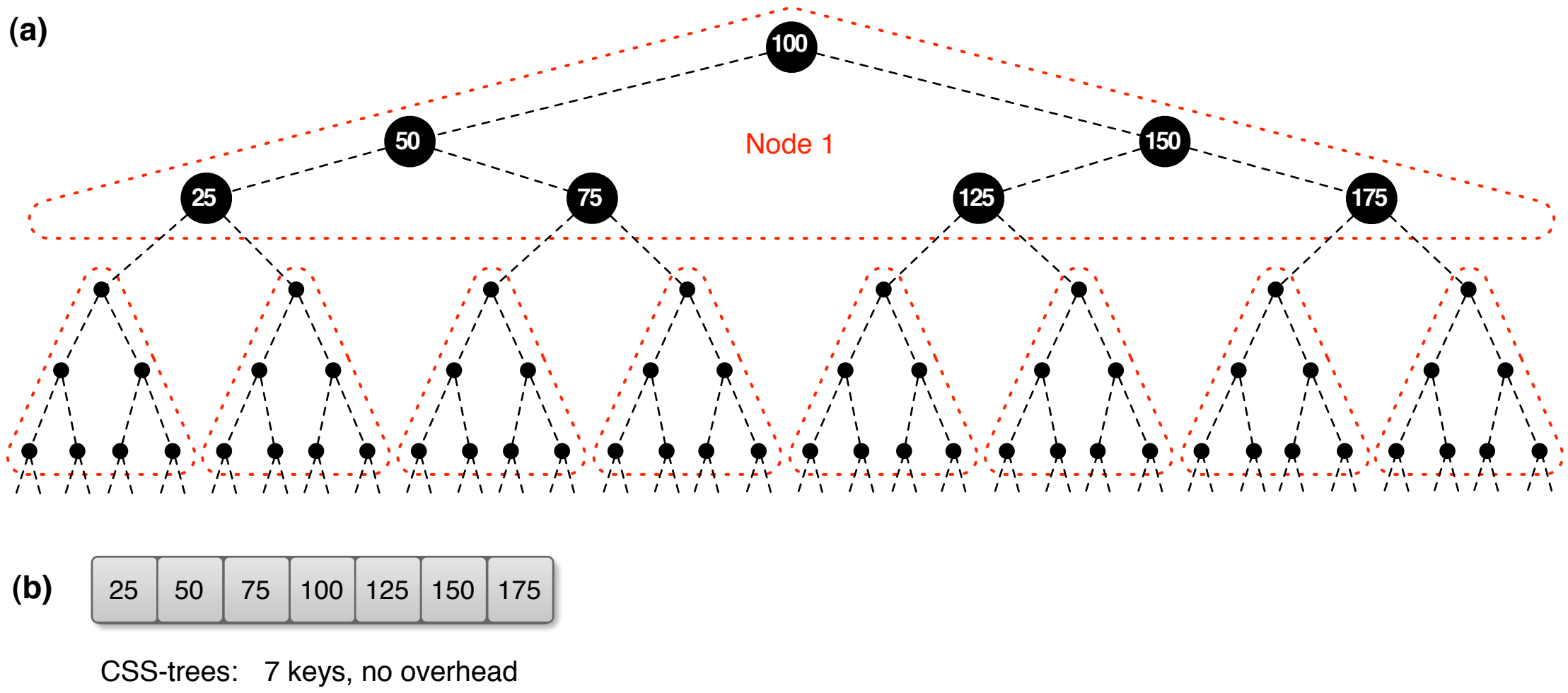

**Figure 3.1:** Logical and physical layout of CSS-trees.
(a) Logical layout: Binary nodes are grouped into nodes (red triangles). References to child nodes are virtual: No pointer is stored.
(b) Physical layout of CSS-tree Node 1: Only the keys are stored.

Index search in CSS-trees works much the same as in $B^+$-trees: The main difference is using node index calculation instead of pointers to traverse the tree. Algorithm 3.1 shows the resulting algorithm.

However, due to the removal of pointers and storing everything in one consecutive array of data, CSS-tree structure is more 'rigid': CSS-trees cannot grow dynamically. Insert, delete, and update operations are handled is by means of a full rebuild of the tree. For usage in online analytical processing where write operations are processed overnight in batches this is viable. But it does mean that special care needs to be exercised when using CSS-trees in write-heavy online transactional processing. Possible improvements are using techniques like differential indexing to batch together updates [SL76].

[2] Assuming 32-bit integer keys and 32-bit child pointers.

```
input: Key to search for key
input: Array of node data nodes
input: Maximum number of node children fanout

nodeIdx ← 0;

while nodeIdx does not refer to a leaf node do
    childIdx ← perform binary range search for key
                in keys of node nodes[nodeIdx];
    nodeIdx ← nodeIdx * fanout + childIdx + 1;
end

keyIdx ← perform binary search for key
          in keys of node nodes[nodeIdx];

if keyIdx ≠ nil then
    return nodes[nodeIdx].values[keyIdx];
else
    return nil;
end
```

**Algorithm 3.1**: An implementation of the CSS-tree search algorithm.

## 3.2 CSB$^+$-Trees

Another way to overcome the inflexibility of CSS-trees is by changing the structure to a hybrid scheme: Cache-sensitive B$^+$-trees (CSB$^+$-trees, [RR00]) are an interesting compromise between CSS-trees and B$^+$-trees.

Figure 3.2 shows the structure of CSB$^+$-trees, which is similar to the structure of CSS- and B$^+$-trees. Solid lines between nodes visualize physical pointers that need to be stored as part of the physical layout. Dashed lines between nodes visualize virtual references to child nodes: The exact address of a child node needs to be calculated based on the index of the child and the pointer to the first child node.

By storing all children of a node together in one contiguous memory area,

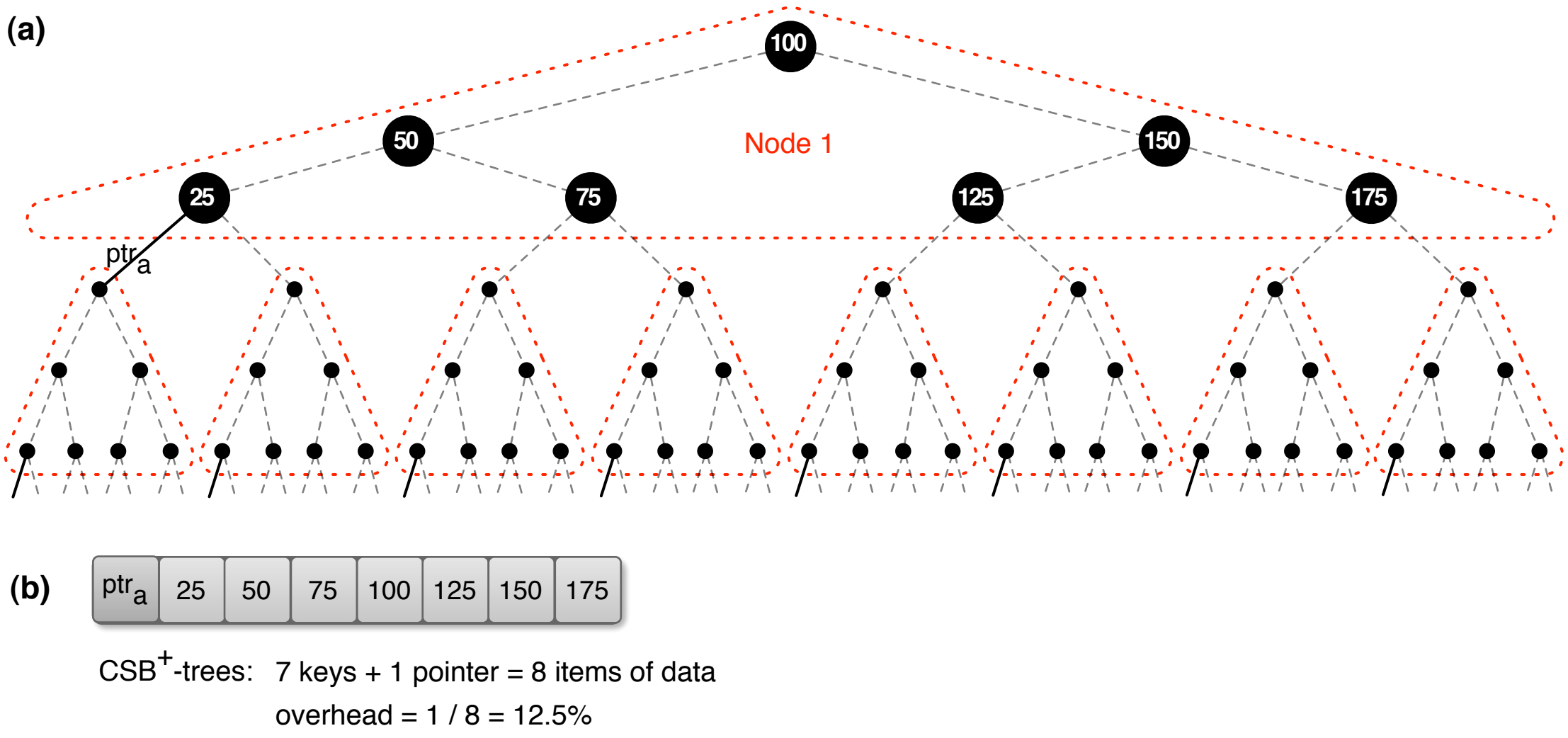

**Figure 3.2:** Logical and physical layout of CSB$^+$-trees.
(a) Logical layout: Binary nodes are grouped into nodes (red triangles). References to all but the first child nodes are virtual: No pointer is stored.
(b) Physical layout of Node 1: Only one pointer is stored per node.

**input**: Key to search for *key*
**input**: Root node of tree *root*

$node \leftarrow root$;

**while** *node* is not a leaf node **do**
    $childIdx \leftarrow$ perform binary range search for *key* in keys of *node*;
    $node \leftarrow *(node.firstChild + childIdx * childSize)$[3];
**end**

$keyIdx \leftarrow$ perform binary search for *key* in keys of *node*;

**if** $keyIdx \neq nil$ **then**
    **return** $node.values[keyIdx]$;
**else**
    **return** *nil*;
**end**

**Algorithm 3.2**: Implementation of the CSB$^+$-tree search algorithm.

they only need to store one pointer for all children. At the same time it is still possible for parts of the tree to grow in size without needing to perform a full rebuild of the CSB$^+$-tree. This means that CSB$^+$-trees can support incremental updates in a fashion similar to B$^+$-trees, but with much lower space and cache utilization overhead than B$^+$-trees, making them a good match for update-frequent workloads, for example in OLTP. In particular, they are a better match than CSS-trees for those workloads.

Algorithm 3.2 shows the implementation of CSB$^+$-tree search. Apart from the arithmetic needed to calculate the address of child nodes, it is identical with B$^+$-tree search.

[3]Note: If this is implemented in C and *node.firstChild* is appropriately typed then adding *n* to that automatically advances *n full* child items – the multiplication by *childSize* would need to be removed. It is just used in the algorithm pseudo-code because it is not C.

## 3.3 k-ary Search

Binary search does only one comparison at a time to split the search domain into two partitions. This is optimal from an asymptotic point of view.

Modern CPUs provide support for SIMD instructions. These allow us to execute multiple parallel operations with one CPU instruction. Currently, they allow us to do four 32-bit integer comparisons in one cycle [Int10].

However, binary search only performs one comparison per cycle. It only utilizes one of four comparisons that the CPU could be doing in one cycle, throwing away computational power by not utilizing it.

The idea behind $k$-ary search [SGL09] is simple: Instead of performing only one comparison and obtaining two partitions, it performs $k$ comparisons and obtains $k+1$ partitions. For current CPU generations the optimal value for $k$ is 4. Figure 3.3 shows a sample run of k-ary search on the same sorted data that binary search was run on in Figure 2.1 on page 15. Binary search needs four steps to completion; k-ary search needs only two.

But there is a big issue with directly running k-ary search on a sorted array: SIMD instructions require all of the data to be present contiguously in one SIMD register. While it is possible to load data that is not contiguous in main memory into a SIMD register, this is very inefficient. It is much faster when all four 32-bit integer numbers, which we want to compare against the search key in one step, are kept contiguously in memory. This is easy to achieve when we reorganize the elements of the key array before running the algorithm. Figure 3.4 shows a run of k-ary search with a reorganized array.

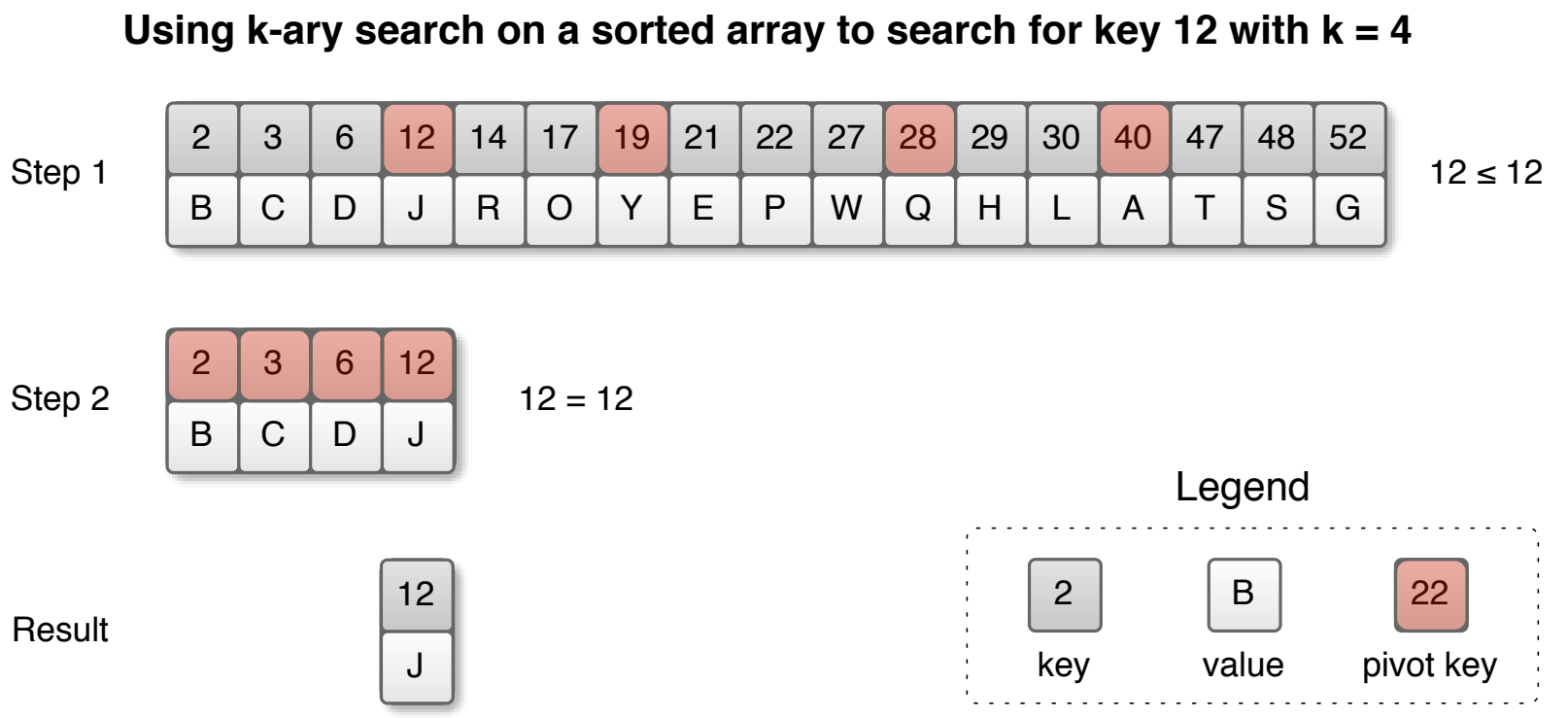

**Figure 3.3:** k-ary search with $k = 4$ decreases search area by 75% after each step until no keys remain to be compared or a matching value is found.

Interestingly, the array we obtain by this reorganization is very similar to the linearized tree representation of CSS search: In fact both have the very same underlying structure; the only difference is in the size of nodes.

Let us briefly summarize existing approaches before moving on:

k-ary search performs better than binary search. Binary search just reads and compares one key per step. k-ary search on a linearized tree representation reads and compares $k$ consecutive keys per step. k-ary search shares one of binary search's problems: What happens when $k$ is smaller than the number of keys that could fit into one cache line?[4] The CPU will fetch a full cache line, but we will only look at a part of it before requesting the next cache line and only using part of it again.

CSS search reorganizes key data into blocks of cache-line-sized nodes. k-ary search reorganizes key data into blocks of SIMD-register-sized nodes. On current hardware, cache line size is four times that of SIMD registers. With these approaches we can either have cache-optimal or SIMD-optimal behavior, but not both. Is there a way of unifying them, obtaining cache- and SIMD-optimal behavior at the same time?

[4]This is already the case for modern CPUs. Current CPUs use cache line sizes of 128 bytes. k-ary search with $k = 4$ and 32-bit integer keys would only read 32 bytes at a time, wasting 75% of the data in a cache line.

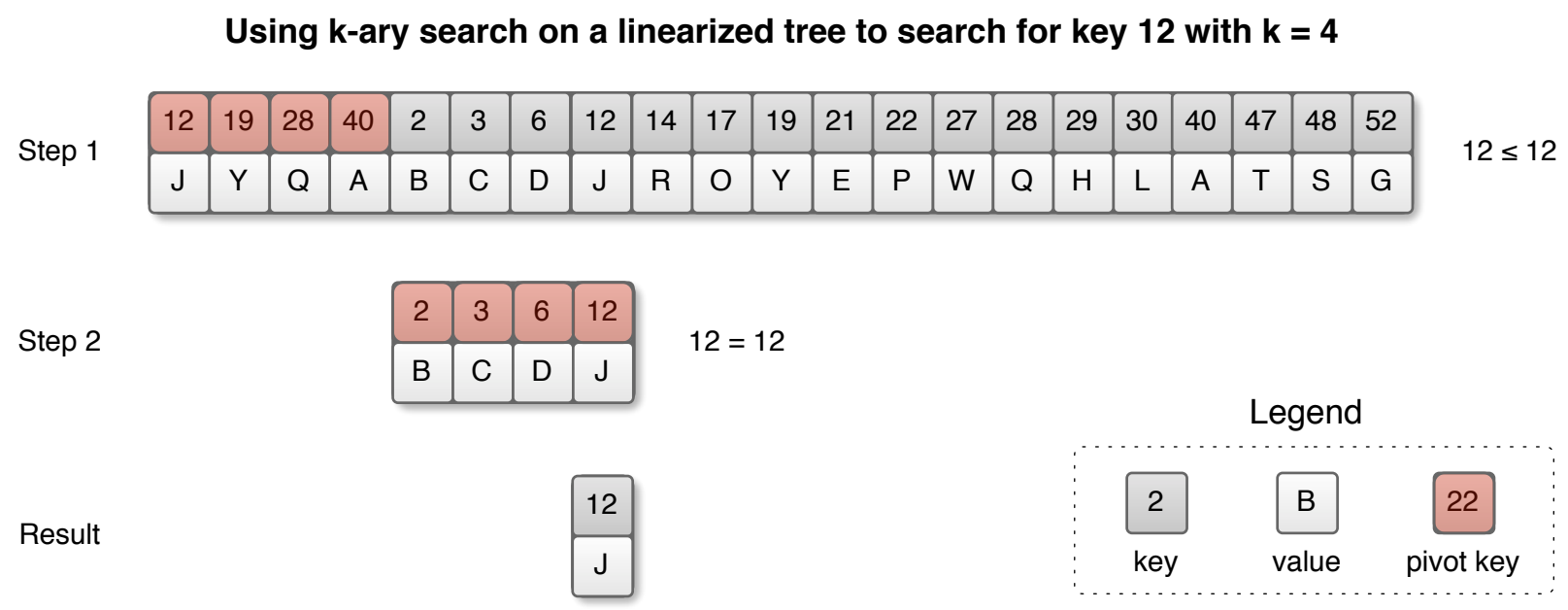

**Figure 3.4:** Reorganizing the key array allows us to use contiguous SIMD register loads, which are much faster than manual scatter-gather loading.

## 3.4 Fast Architecture Sensitive Tree Search

Fast architecture sensitive tree search (FAST, [KCS+10]) unifies the optimality properties of CSS-tree search and k-ary search. Just like those two approaches, it uses a binary tree and organizes it inside of a single contiguous array in such a way that the resulting node access sequence is closely optimized to features of modern hardware architecture. In contrast to those earlier approaches, however, it can actually optimize for multiple hardware architecture features at the *same* time.

The ideas behind FAST are new: They have only recently been published in June at SIGMOD 2010, just three months before the publication of this thesis. Unfortunately, despite requests, the reference implementation is not available. All of the following information is from the paper and our own reimplementation of the ideas presented therein.

In addition to optimizing data layout for SIMD register size and cache line size, FAST also optimizes data layout for page size. Virtual memory management on modern hardware works by using virtual addresses for every memory request. Virtual addresses are split into a page selector and an offset into that page. Before it can actually read the data for a memory request, the CPU needs to look up the physical start address of that page in memory. The mapping of a page selector to a physical page start address is done by lookup in the page table, which is also located in memory. In order to speed up the process of virtual address translation, CPUs have a so called translation lookaside buffer where the physical start addresses of most recently accessed pages are stored[5].

The goal of page-, cache-line-, and SIMD-blocking in FAST is to keep memory requests inside the same memory page, cache line, and SIMD data for as long as possible. Page blocking allows FAST to make optimal use of the translation lookaside buffer, cache line blocking allows FAST to optimally utilize data cache resources, and SIMD blocking allows for using k-ary SIMD search with fast SIMD register loading from contiguous memory.

The way this is achieved is by multi-level reordering of key data: Figure 3.5 shows the structure of nodes inside a single page node employed by FAST.

[5]This was in fact the first cache that became part of CPUs.

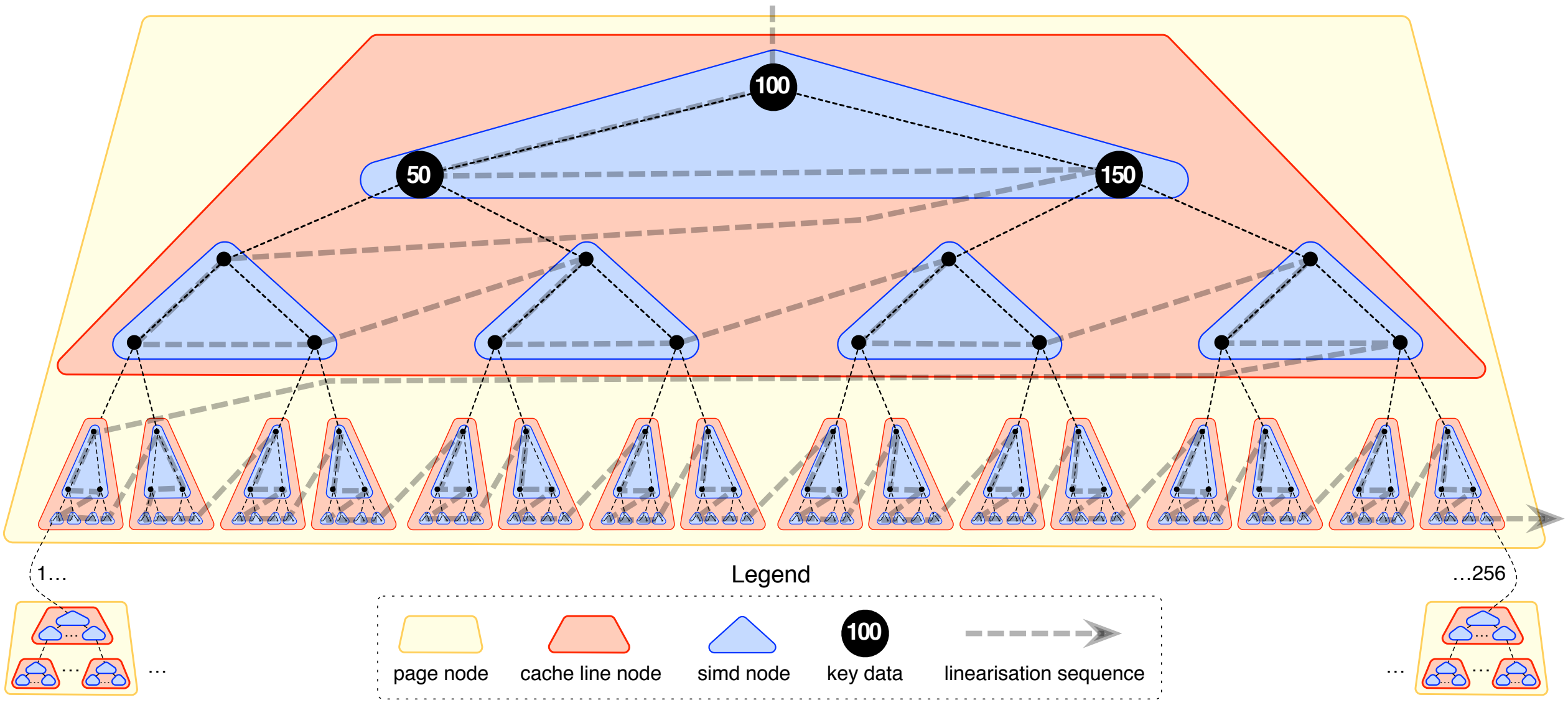

**Figure 3.5:** Intra-page layout of FAST-trees with sketch of inter-page layout.

Page nodes themselves form a tree and do have other page nodes as children, but these are only sketched at the bottom of Figure 3.5. All page nodes have the same structure as the one displayed.

In the figure and the following explanation we decided to use the following parameters for the hierarchical blocking. These are not the optimal parameters from a run-time point of view — they are, however, the ones that make the principles underlying the hierarchical blocking easily comprehensible:

- SIMD nodes are composed of two levels of binary search key nodes. Note that this results in only three keys per SIMD node, and only three comparisons at a time in k-ary SIMD search. Current CPU generations could actually perform four comparisons, however. SIMD nodes have a fanout of four SIMD child nodes. SIMD node size is 12 bytes[6]. This is consistent with the parameters from Kim et al.[7]
- Cache line nodes are composed of two levels of SIMD nodes. This results in a total of five SIMD nodes per cache line node, of which two

[6] Assuming 32-bit integer keys.

[7] Our reimplementation actually supports the general case of an arbitrary number of SIMD keys. If we let the underlying key nodes have a fanout of three, we obtain SIMD nodes with four keys and five SIMD child nodes.

are visited per search. There is a total of fifteen keys per cache line node. Cache line node size is 60 bytes[6]. The typical cache line size of current CPUs is 64 bytes. Cache line nodes have a fanout of sixteen cache line child nodes. This is consistent with the parameters from the paper.

- In this example, page nodes are composed of two levels of cache line nodes. This results in a total of 17 cache line nodes per page node, of which two are visited per search. There is a total of 85 SIMD nodes per page node, of which four are visited per search. This results in a total of 255 keys per page node, and a page node size of 1020 bytes[6]. Page nodes have a fanout of 256 page child nodes. These values are not optimal, but make it easier to explain hierarchical blocking. The original paper uses page node depths of 10, or 19 binary search key nodes[8]. A typical value for page size in current operating systems is 4096 bytes. A page node depth of 10 binary search key nodes would result in a total page node size of 4092 bytes.

Keys are laid out from largest to smallest block units:

- First, the data is split into page nodes. Page nodes are processed level by level, starting from the top of the tree to the bottom of the tree.
- Inside each page node, data is split into cache line nodes. Cache line nodes inside of the current page node are again processed level by level, starting from the top of the page node's subtree to the bottom of the page node's subtree.
- Finally, cache line nodes are split into SIMD nodes. SIMD nodes inside of the current cache line node are processed level by level, starting from the top of the cache line node's subtree to the bottom of the subtree.

The resulting processing order is also visualized by the faded, dashed arrow in Figure 3.5. In search the layout allows us to process data in block units, hierarchically proceeding from the largest to the smallest block unit:

[8]The depth of 19 is only used when the operating system offers support for huge memory pages. This results in even less TLB misses.

1. At the beginning, we find a full page node at the start of the key array. We process this page node.

   (a) At the start of this page node, we find a full cache line node. We process this cache line node.

      i. At the start of this cache line node, we find a full SIMD node. We process this SIMD node. We use k-ary search to find the branch index of the SIMD node child of the current SIMD node that we should next process.

      ii. We process this SIMD node. We use k-ary search to find the index of the appropriate child branch.

      By combining together the two SIMD child branch indices from steps 1(a)i and 1(a)ii, we find the branch index of the cache line node child of the current cache line node that we should next process.

   (b) At the start of this cache line node, we find a full SIMD node...

      i. Process this SIMD node. Find the next SIMD node to process.

      ii. Process this SIMD node. Find the next SIMD child branch.

      By combining together the two SIMD child branch indices from steps 1(b)i and 1(b)ii, we find the next cache line branch index.

   By combining together the two cache line branch indices from steps 1a and 1b, we find the branch index of the page node children of the current page node that we should next process. We have just processed the first page node.

2. We can use the same approach to process the next page node, and the cache line nodes inside that page node, and the SIMD nodes inside that cache line nodes, and so on until we reach the end of the page node hierarchy. We can then combine together the page branch indices, to find the index of the leaf node potentially containing the search key.

3. Inside that leaf node, we can then use regular SIMD search to find the position of the search key, if it exists.

# Chapter 4

# The NitroGen Approach

> "In programming, we deal with two kinds of elements: Procedures and data. (Later we will discover that they are really not so distinct.)"[1] — Abelson & Sussman [AS96]

All of the CPU-optimized index structures we have seen so far have dealt with optimizing the layout of the involved index structures in such a way that the CPU's caches are optimally utilized[2].

In achieving that goal they have focused solely on the data caches and the translation-lookaside buffer of modern CPUs. However, all modern CPUs also come with fast instruction caches for caching an application's code.

Table 4.1 shows the size of the code responsible for performing the index search operations in implementations of select index structures and search algorithms. These numbers ignore code that is not part of the index search itself, such as the code for bulk-loading data into an index etc. These numbers have been obtained by analyzing the executables produced by GNU's C compiler with the `objdump` tool for Linux and `otool` for Mac OS X. They take into account the concrete encoding of machine instructions. The executables

[1] Or, more extremely: "Data is just dumb code, and code is just smart data."

[2] k-ary index search has focused on the orthogonal goal of utilizing non-cache CPU resources. Binary search is not optimized for modern CPUs.

| Algorithm | Tuned for | Size | Remaining Instruction Cache[3] |
|---|---|---|---|
| Binary Search | Performance | 193 byte | 99.4% |
| | Code size | 128 byte | 99.6% |
| CSS-tree Search | Performance | 686 byte | 97.9% |
| | Code size | 348 byte | 98.9% |
| FAST-tree Search | Performance | 1503 byte | 95.4% |
| | Code size | 1503 byte[4] | 95.4% |

**Table 4.1:** Code size for index search implementations: The instruction cache of modern CPUs is largely under-utilized in index search.

have been tuned with flags `-Os` for code size and `-O3` for performance.

As can be seen in Table 4.1 the size of the code used in index lookup operations tends to be much smaller than the available instruction cache size. Instruction cache remains largely under-utilized during index search.

In all of the approaches discussed so far, instruction cache has been completely ignored. Nobody has thought of ways for optimizing index structures to take into account instruction cache.

Instruction cache remains the only CPU cache resource that index structures have not been optimized for so far.

However, there is no way of directly loading data from main memory into or through the instruction cache. It has been designed to be used for caching code. The only way of getting code into the instruction cache is through executing the code.

So one question remains: Is there a way of utilizing the instruction cache for storing the actual data of our index structures when it has only been designed to be used for caching code?

[3] Assuming an instruction cache size of 32 KB, which is the case for current generations of Intel CPUs. Current generations of AMD CPUs use 64 KB of instruction cache.

[4] Telling GCC to tune this code for size actually results in bigger code than telling it to tune for speed, resulting in 2056 bytes of code. This is because the additional inlining it performs when tuning for speed allows it to throw away unused portions of code that it would otherwise have to keep around.

## 4.1 The Fundamental Idea of NitroGen

The NitroGen framework aims to add utilization of instruction caches to the repertoire of existing index structures and algorithms. As it is only possible to store code in the instruction cache, and the only way of getting code there is by executing it, NitroGen achieves its goal by using code generation to transform data into code. It needs to create a copy of the functions used in index search and specialize those copies with the actual key data.

By transforming the top of an index structure into code specialized with concrete data, the top of the index can then be stored in the CPU's instruction cache. We can just execute the dynamically created code to search the top of the tree, and switch back to the generic non-specialized version of the algorithm to search through the remaining data.

Figure 4.1 shows one sample application of NitroGen. To illustrate the concepts by example we use NitroGen to add instruction cache utilization to the simplest possible search algorithm: Linear search on unsorted arrays of key and value data. Figure 4.1a shows the index search code and data for one sample instance of unmodified linear search. Figure 4.1b shows the result of using NitroGen to merge together the index search code with the concrete data for this sample instance of linear search.

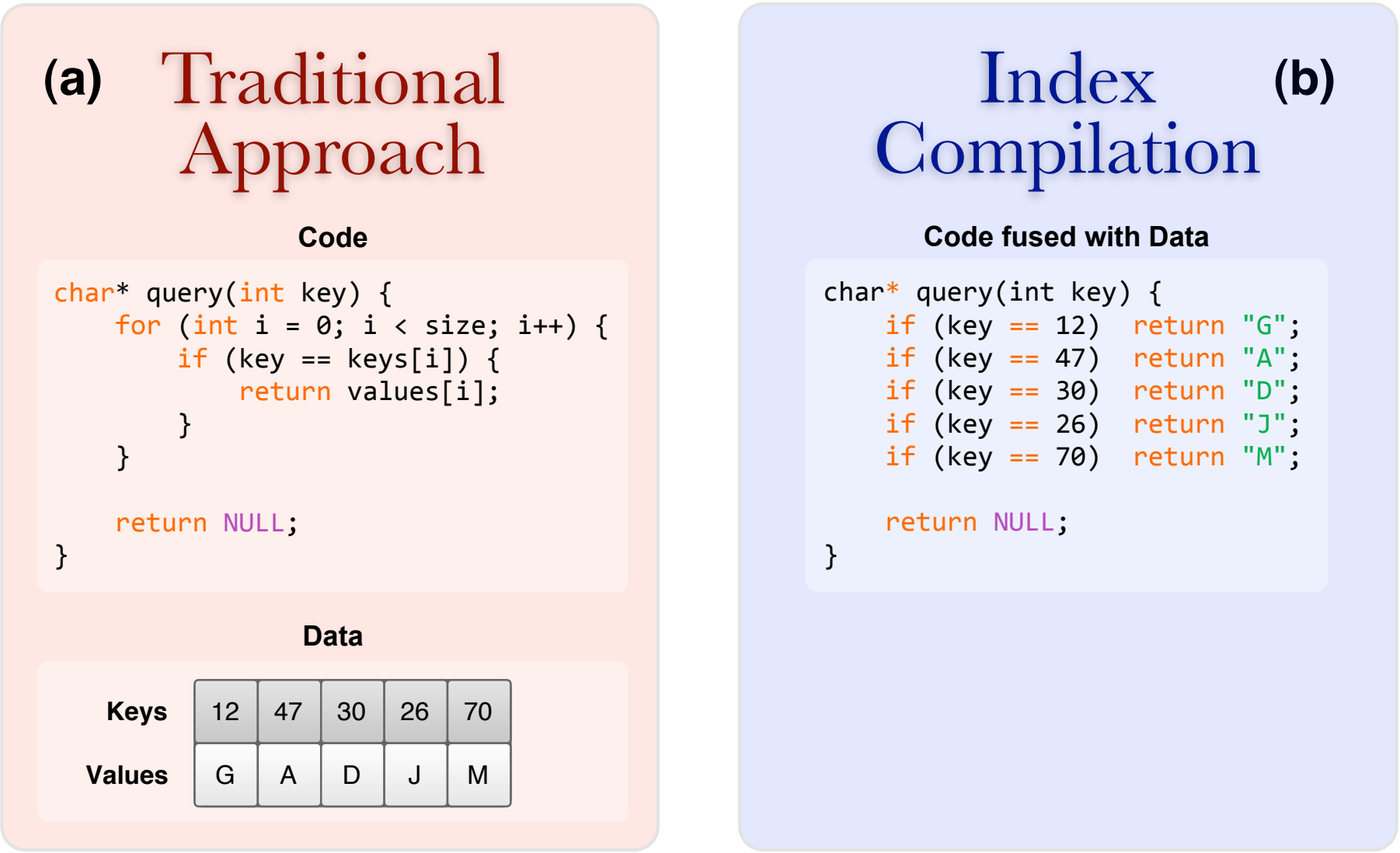

**Figure 4.1:** Sample run of NitroGen for linear search.

Just like CSS- and FAST-trees, NitroGen is heavily optimized for read-intensive workloads, such as the ones commonly seen in OLAP environments. This is because every update triggers run-time code generation. Compared to OLTP index structures like $B^{+}$-trees, updates are more expensive. NitroGen however only converts the top $n$ levels of the index tree into code. The top of an index tree changes less frequently than the lower levels, due to the way node splits propagate up the tree.

Update costs can further be decreased by using JIT run-time code generation libraries to directly generate machine instructions instead of generating C code at run time and invoking a compiler. Due to usage of `jmp` instructions, it is also possible to split the generated code into pieces allocated in separate memory, and only update the affected pieces. This can further reduce the cost of update operations.

In order to obtain the biggest possible performance gain from the available instruction cache, it is imperative to efficiently encode the data into code. The better the encoding, the more data can be fit into the instruction cache.

## 4.2 Current Status

In the following, we briefly describe the current state of our implementation of NitroGen for different base index structures and algorithms. We will start with describing the implementation of NitroGen for binary search, which is reasonably simple.

Our current prototype implementations do not yet employ run-time code generation. To ease initial development and experimental exploration, we decided to use a static upfront code generation scheme implemented in a high level programming language. This allowed us to quickly move through iterations of the implementation process, experimenting with different ways of implementing the conceptual ideas. It requires all of the data to be available at compile time. Our prototype implementations are evaluated in Chapter 5.

### 4.2.1 NitroGen for Binary Search

For binary search it turned out to be viable to make our prototype directly emit C code, and use GCC to generate the resulting machine code instructions. The prototype of the code generator could be implemented in only 270 lines of high level programming language code.

Figure 4.2 shows the code resulting from using NitroGen to fuse a sample instance of binary search together with its data into code. The implementation of binary search that was used is based on Algorithm 2.1 described on page 17. The structure of the resulting code (Figure 4.2a) is very similar to that of the key-value data array (Figure 4.2b).

**(a)**

```
int compiled_bin_search(int key, char** val) {
    int result;              /* Range [0, 7] -------- */

    if (key <= 3) {              /* Range [0, 3] ---- */
        if (key <= 1) {            /* Range [0, 1] -- */
            if (0 == key) { result = "G"; goto done; }
            if (1 == key) { result = "A"; goto done; }
        } else {                   /* Range [2, 3] -- */
            if (2 == key) { result = "D"; goto done; }
            if (3 == key) { result = "F"; goto done; }
        }
    } else {                     /* Range [4, 7] ---- */
        if (key <= 5) {            /* Range [4, 5] -- */
            if (4 == key) { result = "C"; goto done; }
            if (5 == key) { result = "H"; goto done; }
        } else {                   /* Range [6, 7] -- */
            if (6 == key) { result = "E"; goto done; }
            if (7 == key) { result = "B"; goto done; }
        }
    }

    return 0;
done:
    *val = result;
    return 1;
}
```

**(b)**

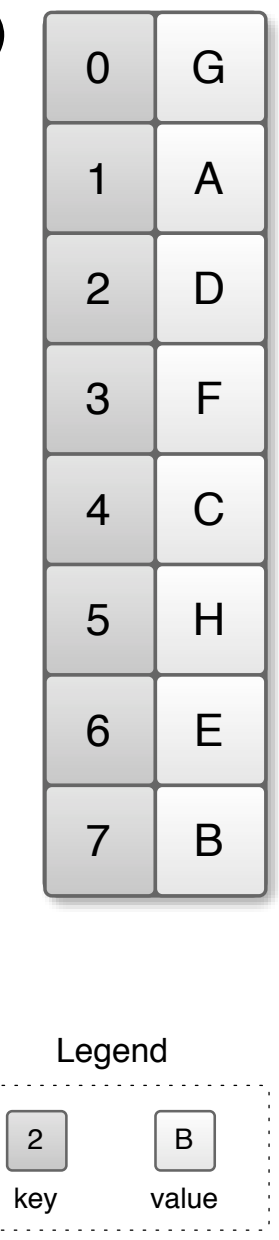

**Figure 4.2:** Sample run of NitroGen for binary search.

### 4.2.2 NitroGen for CSS-Tree Search

Our initial prototype implementation for CSS-tree search also directly generated C code, and used GCC to generate the resulting machine code instructions. However, this soon turned out to be much too slow to carry out any serious experimental evaluation of the performance of the resulting code[5].

So we replaced the prototype implementation emitting C code. Instead we decided to emit assembler code, and use the GNU assembler to turn that into machine code instructions. This allowed us to compile code that took GCC more than 20 hours to compile in less than half an hour.

Figure 4.3 shows the code resulting from using NitroGen to fuse a sample instance of CSS-tree search together with its data into code. The underlying implementation of CSS-tree which was used here is based on

[5]For reasonably sized input data sizes (e.g. 128 MB), GCC took more than 20 hours of time to compile the code. For slightly bigger inputs (but still $< 1$ GB), it was possible for GCC to run out of virtual memory while compiling, despite running on a machine with 10 GB of main memory, and despite turning off compiler optimizations with the `-O0` flag.

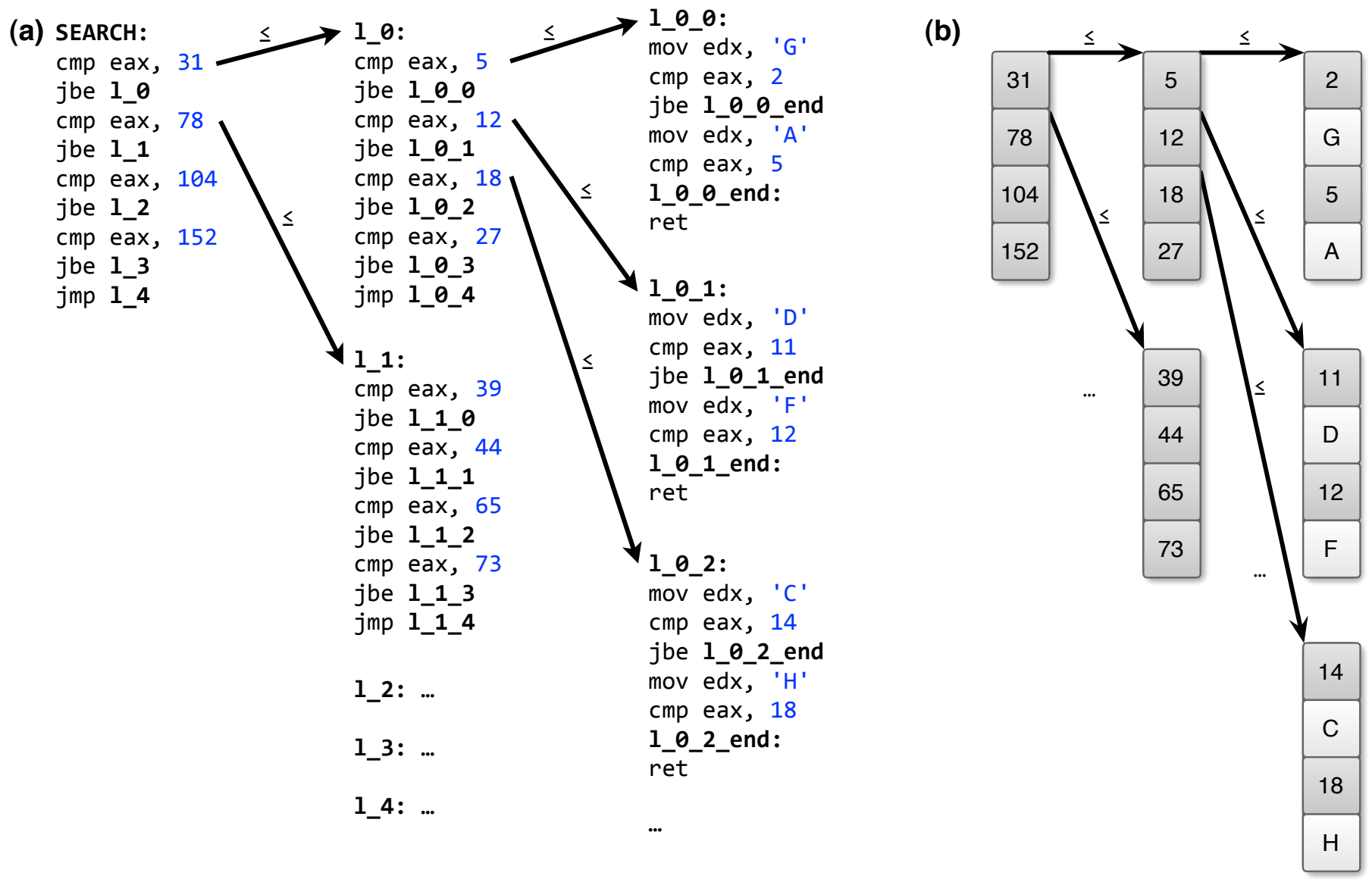

**Figure 4.3:** Sample run of NitroGen for CSS-tree search.

Algorithm 3.1 described on page 24. The structure of the resulting code (Figure 4.3a) is very similar to that of the base CSS-tree (Figure 4.3b).

To achieve good efficiency in encoding data to code, we use some interesting tricks. Figure 4.4 shows the preamble, which is called for each lookup of a key in the NitroGen index-compiled CSS-tree. It then calls the generated SEARCH code from Figure 4.3 which encodes the actual data. By intricate setup of the preamble, we can reduce the size of the generated search code. Let us examine some of the optimizations:

```
_compiled_css_search:
push rbp
mov rbp, rsp

; move key argument to eax:
; cmp reg defaults to eax -> less bytes
mov eax, edi
; perform search
call SEARCH

; if key not found, return value will be 0
mov eax, 0
; flag registers still set from SEARCH
jne EXIT

; if key found, return value will be 1
mov eax, 1
; and need to store value in value pointer
mov [rsi], ecx

EXIT:
leave
ret
```

**Figure 4.4:** Preamble of NitroGen CSS-tree search code: Careful setup allows for a more efficient encoding of the generated search code.

1. The preamble uses `call` to invoke the search code: This lets the `SEARCH` use `ret` (instead of `jmp`) in leaf nodes to end the search. A `ret` instruction can be encoded as 1 byte, a `jmp` needs 3 bytes if the target address is nearby. Otherwise it needs 5 bytes. In this case we would need to jump back to the very beginning of the preamble, which would not be nearby. We save four out of five bytes, or 80%.

2. The checks for whether the key was found or not, and the logic for handling those cases is entirely contained in the preamble. The search code in leaf nodes just keeps comparing keys until it reaches the end of the leaf node. In case it finds a key larger than or equal to the requested search key, it directly jumps to the end of the leaf node.

On every comparison (done with `cmp`) the CPU updates the so called flag register which contains the result of the last comparison. The contents of the flag register are set to the result of the last comparison when returning from the search code back to the preamble. At the end of the leaf node, the last comparison will have been against the first key $\geq$ the search key. The preamble can then use the flag register to check whether the last compared key is equal to the search key, and do the right thing. This allows us to save more than 20 bytes per leaf node, equivalent to around two key-value pairs.

3. By storing the search key in register `eax` we can use a more efficient encoding for the `cmp` instructions, saving 1 byte per comparison.

4. The search code unconditionally moves the value into a CPU register, without checking whether it actually belongs to the search key. Wrong values just get overridden later on. At the very end, the preamble can check whether the last written value corresponds to the search key by checking the CPU flag register. This allows us to save one conditional jump per pair of key-value data, which is equivalent to 4 bytes.

Assuming a leaf node size of 64 key-value-pairs, this results in roughly 14 bytes to store one single key-value pair as code. Assuming an internal node size of 32 keys, this results in roughly 10 bytes to store one key as code. There is still potential for optimizing the encoding, for example by introducing k-ary SIMD comparison.

## 4.3 Future Work

NitroGen shows potential for improving existing index structures and algorithms. We next intend to implement a variant of it for FAST-tree search, making use of k-ary SIMD comparison, and adding optimizations for the page size employed by the operating system.

# Chapter 5

# Experimental Evaluation

This chapter is concerned with evaluating the performance of index structures and search algorithms on concrete modern hardware systems.

We will have an in-depth look at the configuration of the hardware used to run the experiments, examine the parameter space of index search algorithms, pick a subset of index search algorithms to evaluate in detail, pick experiments to run, and present the results of running them, all in the hope of answering some basic questions on index search algorithms:

- What are the optimal parameters of the index search approaches we discussed in Chapters 2 and 3? What is the difference between optimal parameters for different platforms?
- How do these index search approaches perform when parametrized with those optimal parameters on concrete instances of modern hardware?
- Is it possible to improve these base approaches with index compilation?
- How do we pick the parameters of index compilation?
- What can be gained by applying index compilation to these base approaches?
- What is the effect of the node size on index search approaches?
- What is the effect of key skew on index response times?

## 5.1 Experimental Setup

For our evaluation of index search algorithms, we used two differently configured hardware systems, whose most important properties are described in Table 5.1. System A is a Intel Xeon system utilizing the Xeon E5430 CPU, which was released at the end of 2007. System B is an AMD Opteron system with an Opteron 2378, released at the end of 2008. Both are running SUSE Linux with recent versions of the Linux kernel and GCC. System A has a slightly larger amount of main memory than System B (10 GB vs. 8 GB). System A has a slightly higher CPU clock than A (2.6 GHz vs. 2.4 GHz). It also has a larger amount of second and third level CPU cache (12 MB vs. 8 MB). In contrast, System B has more Level 1 cache than System A (128 KB vs. 64 KB in total). Both use the same cache line and SIMD register sizes.

There is a large space of algorithms and parameters in index search.

| **Property** | **System A** | **System B** |
|---|---|---|
| CPU Vendor & Model: | Intel Xeon E5430 | AMD Opteron 2378 |
| CPU Architecture: | Harpertown (45 nm) | Shanghai (45 nm) |
| CPU Release Date: | November 11, 2007 | November 13, 2008 |
| CPU Frequency: | 2.6 GHz | 2.4 GHz |
| CPU Cores: | 4 cores | 4 cores |
| Main Memory: | 10 GB | 8 GB |
| L1 Data Cache: | 32 KB | 64 KB |
| L1 Instruction Cache: | 32 KB | 64 KB |
| L2 Unified Cache: | 12 MB | 2 MB |
| L3 Unified Cache: | None | 6 MB |
| Cache Line Size: | 64 byte | 64 byte |
| TLB Size: | 512 4 KB pages | 1024 4 KB pages |
| SIMD Register Size: | 128 bytes per register | 128 bytes per register |
| Operating System: | SUSE Linux 2.6.34.7 | SUSE Linux 2.6.34.12 |
| Compiler: | GCC 4.5.0 | GCC 4.5.0 |

**Table 5.1:** Configuration of hardware systems used in experiments.

| Approach | Parameters |
|---|---|
| Common: | • Key element size<br>• Pointer size<br>• Workload parameters[1]<br>• Hardware parameters[2] |
| Binary Search: | • Implementation parameters[3] |
| $B^+$-Tree and B-Tree Search: | • Keys per internal node, and resulting fanout<br>• Internal node layout<br>• Binary search strategy for internal nodes<br>• Key-value pairs per leaf node<br>• Leaf node layout<br>• Binary search strategy for leaf nodes |

**Table 5.2:** List of traditional search algorithms and their parameters.

Table 5.2 lists the parameters for the index search algorithms which we have seen in Chapter 2. Even for a relatively simple algorithm such as $B^+$-tree search we have to pick optimal values for internal node size, leaf node size, if and how to arrange keys in leaf nodes, and what binary search strategies to use in leaf and internal nodes.

For binary search we are going to use a straightforward iterative C implementation based on Algorithm 2.1 from page 17, but with one significant change: We are going to use a cutoff, and switch to linear search when there is less than a specific amount of elements remaining in the key range. Our experiments confirmed the expectation of this being faster than running regular binary search to completion.

We are not going to evaluate $B^+$-tree search, nor B-tree search.

[1]Such as the size of the input data, the number of queries, the key distribution in queries, other access pattern parameters of queries, and the input key distribution.

[2]Especially page size, cache line size, SIMD register size, CPU clock, memory clock, and memory size, but see Table 5.1 for a more complete list.

[3]This includes questions such as these: Does the implementation use iteration or recursion? At what number of remaining elements do we switch over to linear search? Is the case of key equality handled for each comparison, or when the search range has been shrunk to one element? See [Knu98] for a good overview of such details.

| Approach | Parameters |
|---|---|
| CSS-Tree Search: | Same as $B^+$-Tree Search |
| $CSB^+$-Tree Search: | Same as CSS-Tree Search |
| k-ary Search: | • Choice of $k$<br>• Alignment of keys<br>• Whether to use tree layout or discontinuous loads<br>• Implementation parameters similar to binary search |
| FAST-Tree Search: | • Keys per SIMD node, and resulting fanout<br>• Parameters of k-ary search in SIMD nodes<br>• Depth of cache line nodes in SIMD nodes<br>• Depth of page nodes in SIMD nodes<br>• Key-value pairs per leaf node<br>• Leaf node layout<br>• Search strategy for leaf nodes<br>• Whether to use SIMD search in internal nodes<br>• Whether to perform cache line blocking<br>• Whether to perform page blocking |
| NitroGen: | • Number of index-compiled internal node levels<br>• Code layout for internal nodes<br>• Code layout for leaf nodes<br>• Choice of base approach<br>• Parameters of base approach in index-compiled levels[4]<br>• Parameters of base approach in non-compiled levels |

**Table 5.3:** List of CPU-optimized search algorithms and their parameters.

Table 5.2 lists the parameters for the more complex CPU-optimized index search algorithms from Chapters 3 and 4. There is many parameters involved in optimally configuring these approaches.

To ensure a fair comparison of the approaches, we will tune the parameters of the algorithms for the hardware systems from Table 5.1.

[4]It makes sense for these to picked independently of the base approach parameters in non-compiled levels as the optimal values can be quite different.

For the evaluation of CSS-trees we are using the implementation from the original paper by Rao et al. [RR99], which has been published as a part of their implementation of $CSB^+$-trees [RR00]. Out of those two approaches, we are only going to evaluate CSS-trees, due to our focus on OLAP and to the paper on $CSB^+$-trees measuring the read performance of $CSB^+$-trees to be quite similar to one of CSS-trees.

We decided not to evaluate the non-tree-based form of k-ary search. We consider the tree-based form to be more interesting as it does not require the use of SIMD register loading from discontiguous memory access which is quite slow in current CPU hardware generations [SGL09]. We only evaluate k-ary search as part of our reimplementation of FAST-tree search.

We would have liked to use the original implementation of FAST-tree search as described by Kim et al. [$KCS^+10$], and have tried to obtain a copy of their implementation. Unfortunately, despite requests, the reference implementation is not available.

We have implemented index-compilation as part of the NitroGen framework for binary search, and CSS-trees. We are currently working on also implementing index-compilation for FAST-tree search, and intend to make this available as a part of the NitroGen framework.

To summarize, we are going to evaluate binary search, CSS-tree search, FAST-tree search, index-compiled binary search, and index-compiled CSS-tree search. For FAST-tree search we are going to use a custom reimplementation based on the ideas from the paper by Kim et al. [$KCS^+10$]. We use the original implementation of CSS-trees [RR99], and a tuned implementation of binary search as the base lines.

We are going to run the following experiments: An experiment measuring the effect of NitroGen on the performance of CSS-tree and binary search with random and skewed access patterns, an experiment measuring the effect of the keys per internal node has on the performance of index search for 32 MB of input data, and an experiment measuring the performance of the first version our reimplementation of FAST.

## 5.2 Experimental Results

### 5.2.1 Binary Search, CSS-Tree Search, and NitroGen

Figure 5.1 shows the performance of CSS-tree search, as well as binary search, both with and without NitroGen index compilation. There is two copies of the lines: The top group of lines represents uniformly random key references in index lookups. The bottom group obtains more realistic key access patterns, by using frequencies modeled after a Zipf distribution. The gain of index compilation on binary search can be a performance improvement of up to 33%. With this first prototype, we obtain a performance improvement of 6–10% for CSS-tree search, depending on input size and access pattern.

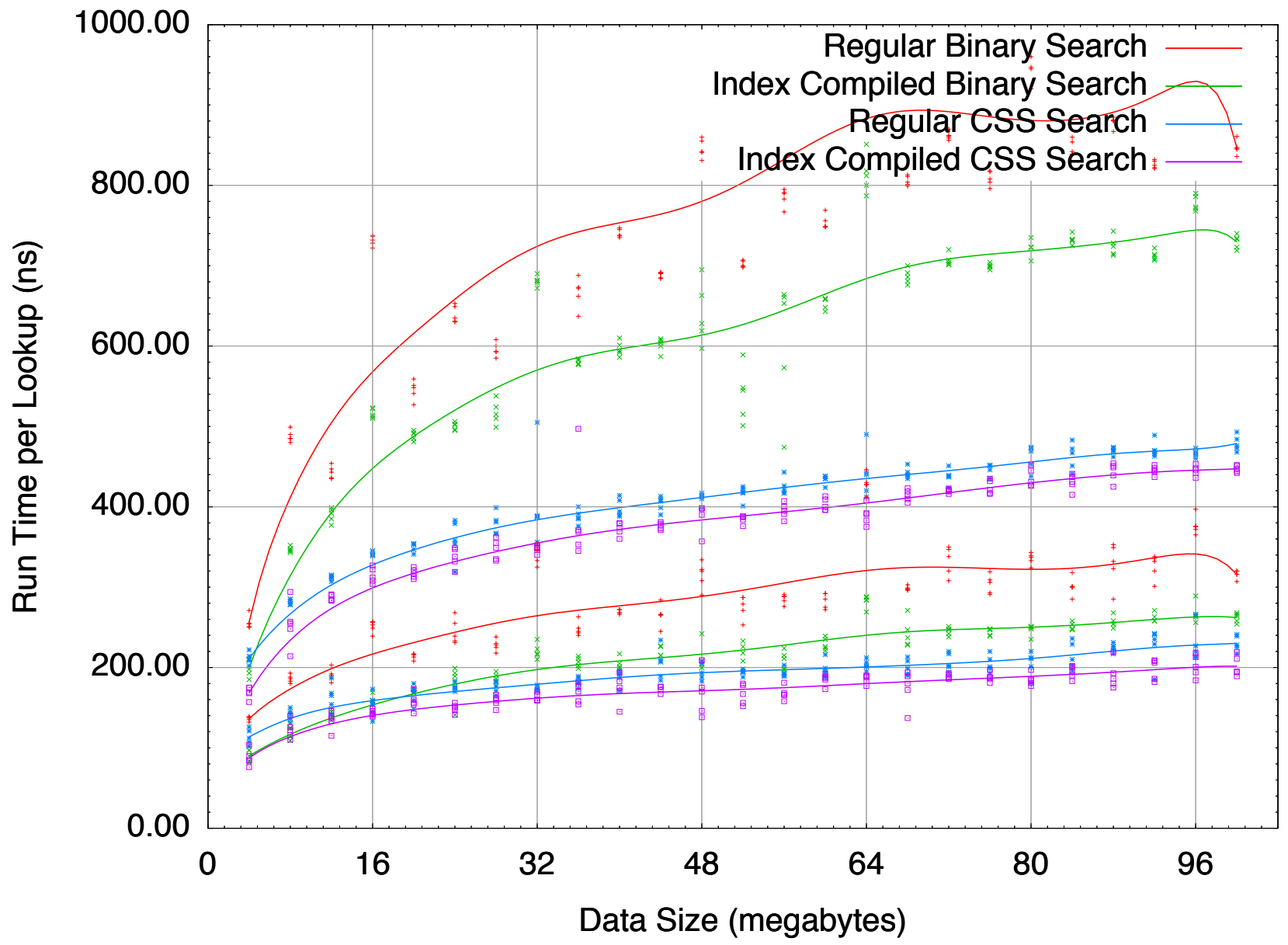

**Figure 5.1:** The performance of CSS-tree search and binary search, both with and without NitroGen. The top four lines represent random key access patterns. The bottom four lines represent key access pattern with key frequencies modeled after Zipf distribution.

### 5.2.2 Effect of Internal Node Size in CSS-Trees

Figure 5.2 shows the effect of the keys per internal node parameter on performance in CSS-tree search for 32 MB of input data as measured on System A.

Whereas the optimal value for CSS-tree search is 32 keys per internal node (equivalent to two cache lines), the optimal value for NitroGen CSS-tree search is 16 keys per internal node. This is consistent with the overhead of turning internal node key data into code.

In this case NitroGen CSS search outperforms regular CSS search by 5% for 16 keys per node.

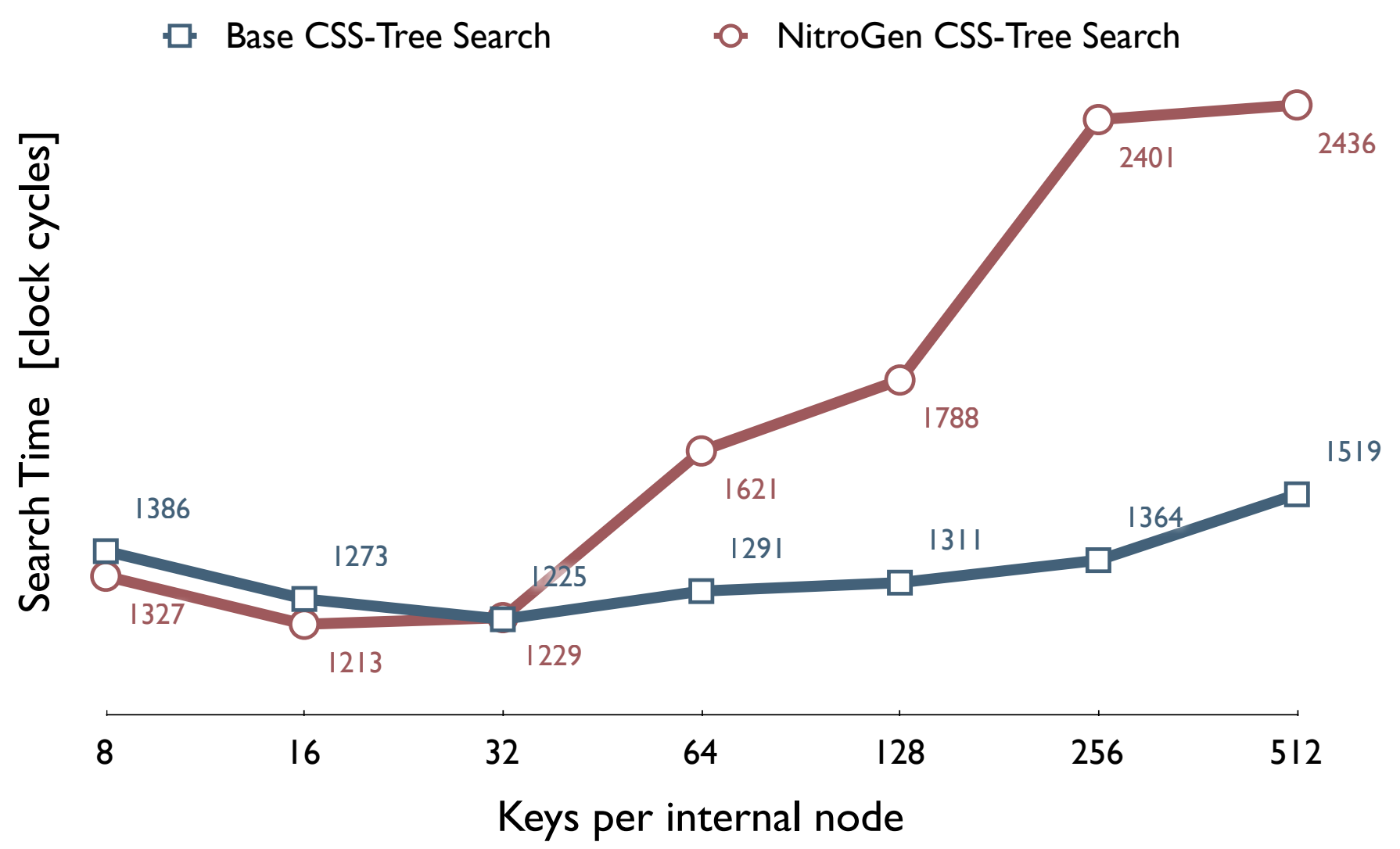

**Figure 5.2:** The effect of the keys per internal node parameter on performance in CSS-tree search for 32 MB of input data as measured on System A.

### 5.2.3 Performance of FAST reimplementation

Figure 5.3 compares the performance of our reimplementation of FAST against the performance numbers reported in the original paper. The original graph reports the relative performance gain obtained by individual features. We also report absolute cycle numbers. These absolute numbers are comparable to those of the original paper. The relative performance gains due to individual features differ. This could be due to a variety of reasons, for example due to different choice of implementation details, hardware architectural differences, or suboptimal algorithmic tuning. Without access to the reference implementation, it is impossible to tell with certainty.

It will be interesting to see how performance is affected after incomplete nodes are handled in the same way in both of the implementations.

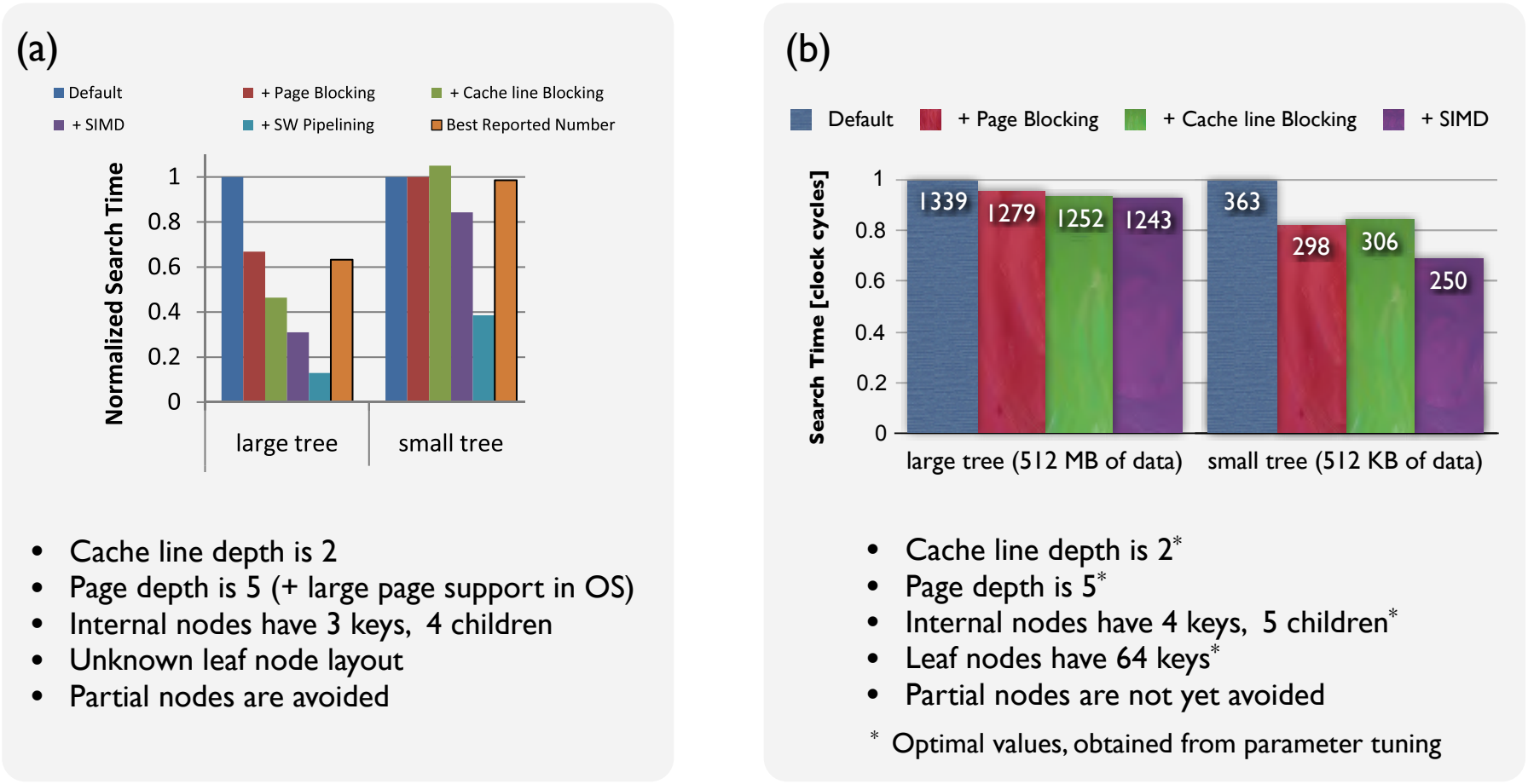

**Figure 5.3:** Performance of FAST reimplementation: (a) Performance as measured in experiments by Kim et al. [KCS$^+$10]
(b) Performance of our reimplementation

# Chapter 6

# Conclusion

## 6.1 Statement of Results

We have discussed the backgrounds and advantages of a wide array of index structures and search algorithms. We have evaluated a representative subset against each other, and discussed the merits of a novel idea: The use of code generation to speed up index structures. We have engineered a prototype implementation, NitroGen, of index compilation for two widely-deployed index search algorithms, and obtained encouraging initial results.

There is ongoing work on a reimplementation of the state of the art index search algorithm, FAST, and ways of utilizing index compilation to make it utilize the instruction caches of modern CPUs, and, ideally, make it even faster.

## 6.2 Statement of Problems left unresolved

Due to the timing of the publication of FAST, we have only had very limited time to discuss some of the results we obtained. It would be interesting to come up with a model for predicting the performance impact of index compilation based on hardware events, such as cache misses. There is ongoing work on reimplementing a regular and an index-compiled version of FAST.

Additionally, there seems to be a wide array of opportunities for follow-up work in the area of index structures, such as the effects of fusing together multiple index structures to obtain best of both worlds hybrid index structures optimized for both OLAP and OLTP. Is it viable to use a read-optimized index structure to represent the root levels of an index, which are less likely to change, while using one supporting inexpensive update operations for the lower levels of the same index, which are more likely to change?

# Bibliography

[ADHW99] Anastassia Ailamaki, David J. DeWitt, Mark D. Hill, and David A. Wood. DBMSs on a modern processor: Where does time go? In *VLDB*, pages 266–277, 1999.

[AS96] Harold Abelson and Gerald Jay Sussman. *Structure and Interpretation of Computer Programs*. MIT Press, Cambridge, second edition, 1996.

[BC07] Stefan Büttcher and Charles L. A. Clarke. Index compression is good, especially for random access. In *CIKM*, pages 761–770. ACM, 2007.

[BM72] Rudolf Bayer and Edward M. McCreight. Organization and maintenance of large ordered indexes. *Acta Informatica*, pages 245–262, 1972.

[BVZB05] Marc Berndl, Benjamin Vitale, Mathew Zaleski, and Angela Demke Brown. Context threading: A flexible and efficient dispatch technique for virtual machine interpreters. In *CGO*, pages 15–26. IEEE Computer Society, 2005.

[Cod70] Edgar F. Codd. A relational model of data for large shared data banks. *CACM*, 13(6):387, 1970.

[Com79] Douglas Comer. Ubiquitous B-tree. *ACM Computing Surveys*, 11(2):121–137, 1979.

[Dua08] Gustavo Duarte. What your computer does while you wait. http://duartes.org/gustavo/blog/post/what-your-computer, 2008.

[HAMS08] Stavros Harizopoulos, Daniel J. Abadi, Samuel Madden, and Michael Stonebraker. OLTP through the looking glass, and what we found there. In *SIGMOD*, pages 981–992. ACM, 2008.

[HP03] Richard A. Hankins and Jignesh M. Patel. Effect of node size on the performance of cache-conscious B+-trees. In *SIGMETRICS*, page 294. ACM, 2003.

[Int10] Intel Corporation. *Intel 64 and IA-32 Architectures Software Developer's Manual.* 2010.

[Kam10] Poul-Henning Kamp. You're doing it wrong. *CACM*, 53(7):55–59, 2010.

[KCS+10] Changkyu Kim, Jatin Chhugani, Nadathur Satish, Eric Sedlar, Anthony D. Nguyen, Tim Kaldewey, Victor W. Lee, Scott A. Brandt, and Pradeep Dubey. FAST: fast architecture sensitive tree search on modern CPUs and GPUs. In *SIGMOD*, pages 339–350. ACM, 2010.

[Knu98] Donald E. Knuth. *The Art of Computer Programming*, volume 3: Sorting and Searching. Addison Wesley, 1998.

[Mau46] John Mauchly. Theory and techniques for the design of electronic digital computers. 1946.

[MBK00] Stefan Manegold, Peter A. Boncz, and Martin L. Kersten. Optimizing database architecture for the new bottleneck: memory access. In *VLDB*, pages 231–246. Springer, 2000.

[McC09] John C. McCallum. Memory Prices from 1957 to 2010. http://www.jcmit.com/memoryprice.htm, July 2009.

[MKB09] Stefan Manegold, Martin L. Kersten, and Peter Boncz. Database architecture evolution: mammals flourished long before dinosaurs became extinct. *VLDB*, pages 1648–1653, 2009.

[Ros07] K.A. Ross. Efficient hash probes on modern processors. In *Data Engineering, 2007. ICDE 2007. IEEE 23rd International Conference on*, pages 1297–1301. IEEE, 2007.

[RPML06] Jun Rao, Hamid Pirahesh, C. Mohan, and Guy Lohman. Compiled query execution engine using JVM. In *ICDE*, pages 23–35, 2006.

[RR99] Jun Rao and Kenneth A. Ross. Cache Conscious Indexing for Decision-Support in Main Memory. In *VLDB*, pages 78–89, 1999.

[RR00] Jun Rao and Kenneth A. Ross. Making B+-trees cache conscious in main memory. In *SIGMOD*, pages 475–486. ACM, 2000.

[SGL09] Benjamin Schlegel, Rainer Gemulla, and Wolfgang Lehner. k-ary search on modern processors. In *DaMoN*, pages 52–60. ACM, 2009.

[SL76] Dennis G. Severance and Guy M. Lohman. Differential files: their application to the maintenance of large databases. *ACM Transactions on Database Systems*, 1(3):256–267, 1976.

[SMA+07] Michael Stonebraker, Samuel Madden, Daniel J. Abadi, Stavros Harizopoulos, Nabil Hachem, and Pat Helland. The end of an architectural era (it's time for a complete rewrite). In *VLDB*, pages 1150–1160, 2007.

[Win10] Markus Winand. Use the index, luke! — a guide to database performance. http://use-the-index-luke.com/, 2010.

[ZR02] Jingren Zhou and Kenneth A. Ross. Implementing database operations using SIMD instructions. In *SIGMOD*, pages 145–156, New York, NY, USA, 2002. ACM.

[ZR04] Jingren Zhou and Kenneth A. Ross. Buffering database operations for enhanced instruction cache performance. In *SIGMOD*, pages 191–202. ACM, 2004.